\documentclass[aps,prresearch,reprint,preprintnumbers,twocolumn,
superscriptaddress,
amsmath,amssymb,aps]{revtex4-2}
\usepackage[utf8]{inputenc}
\usepackage{mathrsfs}

\usepackage{array}
\usepackage{ragged2e}

\usepackage{graphicx}
\graphicspath{{./figs/}}
\usepackage{dcolumn}
\usepackage{bm}

\usepackage{mathrsfs}

\usepackage[normalem]{ulem}

\usepackage{enumitem}

\usepackage{bm}

\usepackage{color}
\usepackage[dvipsnames, table]{xcolor}
\definecolor{blue}{rgb}{0,0,0.5}
\definecolor{lightgray}{gray}{0.95} 

\usepackage[colorlinks,linkcolor=blue,urlcolor=blue,citecolor=blue]{hyperref}
\usepackage{graphicx}
\usepackage{slashed}
\usepackage{enumitem}
\usepackage{bm}
\usepackage{physics}
\usepackage{xspace}
\usepackage[T1]{fontenc}
\usepackage{lmodern}
   
\usepackage{tabularx} 
\usepackage{caption}  
\usepackage{booktabs}  
\usepackage{diagbox}
\usepackage{multirow}

\newcommand{\be}{\begin{equation}}
\newcommand{\ee}{\end{equation}}
\newcommand{\bea}{\begin{eqnarray}}
\newcommand{\eea}{\end{eqnarray}}

\renewcommand{\>}{\rangle}
\newcommand{\mc}{\mathcal}

\newcommand{\noi}{\noindent}
\def\mcB{\mc B}

\def\dof{{\rm dof}}
\newcommand{\MeV}{{\rm MeV}}
\newcommand{\GeV}{{\rm GeV}}

\newcommand{\SM}{{\rm SM}}
\newcommand{\BR}{{\mc B}}

\def\BpKpvv{B^+ \to K^+ \nu \bar \nu}
\def\BpKpa{B^+ \to K^+ a}
\def\refeq#1{Eq.~(\ref{#1})}
\def\refeqs#1#2{Eqs.~(\ref{#1})-(\ref{#2})}
\def\reftab#1{Table~\ref{#1}}
\def\reffig#1{Fig.~\ref{#1}}

\def\refapp#1{Appendix~\ref{#1}}
\newcommand{\BKa}{\BpKpa}
\newcommand{\mua}{\mc B_a}

\def\eps{\epsilon}

\def\nunu{\nu \bar \nu}
\def\de{\partial}

\def\kV#1{(k_V)_{#1}}
\def\kA#1{(k_A)_{#1}}

\newcommand{\bkg}{{\rm bkg}}
\newcommand{\hst}{{\rm hst}}
\newcommand{\rec}{{\rm rec}}
\newcommand{\bin}{{\rm bin}}
\newcommand{\obs}{{\rm obs}}

\renewcommand{\th}{{\rm th}}
\def\ITA{{\rm ITA}}
\def\HTA{{\rm HTA}}
\newcommand{\FVsb}{(F_V)_{sb}}

\newcommand{\camalich}{MartinCamalich:2020dfe}

\newcommand{\georgi}{Georgi:1986df}
\newcommand{\diluzio}{DiLuzio:2020wdo}

\newcommand{\pdg}{ParticleDataGroup:2024cfk}

\newcommand{\BelleIIevi}{Belle-II:2023esi}
\newcommand{\fridell}{Fridell:2023ssf}
\newcommand{\boltonI}{Bolton:2024egx}
\newcommand{\boltonII}{Bolton:2025fsq}

\newcommand{\mcL}{\mc L}

\newcommand{\beal}{\begin{aligned}}
\newcommand{\eeal}{\end{aligned}}

\makeatletter
\g@addto@macro\bfseries{\boldmath}
\makeatother

\makeatletter
\def\p@subsection{}
\makeatother

\newlist{todolist}{itemize}{2}
\setlist[todolist]{label=$\square$}
\usepackage{pifont}
\DeclareOldFontCommand{\rm}{\normalfont\rmfamily}{\mathrm}
\DeclareOldFontCommand{\sf}{\normalfont\sffamily}{\mathsf}
\DeclareOldFontCommand{\tt}{\normalfont\ttfamily}{\mathtt}
\DeclareOldFontCommand{\bf}{\normalfont\bfseries}{\mathbf}
\DeclareOldFontCommand{\it}{\normalfont\itshape}{\mathit}
\DeclareOldFontCommand{\sl}{\normalfont\slshape}{\@nomath\sl}
\DeclareOldFontCommand{\sc}{\normalfont\scshape}{\@nomath\sc}

\newcommand{\bof}[1]{\smallskip \noindent {\bfseries #1}---}

\begin{document}

\preprint{LAPTH-036/25}

\title{The $B^+ \to K^+ \nu \bar \nu$ decay as a search for the QCD axion}

\author{Merna Abumusabh}

\email{merna.abumusabh@iphc.cnrs.fr}

\affiliation{%
{\itshape IPHC, Universit\'{e} de Strasbourg et CNRS, 67200 Strasbourg, France}
}%

\author{Giulio Dujany}

\email{giulio.dujany@iphc.cnrs.fr}

\affiliation{%
{\itshape IPHC, Universit\'{e} de Strasbourg et CNRS, 67200 Strasbourg, France}
}%

\author{Diego Guadagnoli}

\email{diego.guadagnoli@lapth.cnrs.fr}

\affiliation{%
{\itshape LAPTh, Universit\'{e} Savoie Mont-Blanc et CNRS, 74941 Annecy, France}
}%

\author{Axel Iohner}

\email{axel.iohner@lapth.cnrs.fr}

\affiliation{%
{\itshape LAPTh, Universit\'{e} Savoie Mont-Blanc et CNRS, 74941 Annecy, France}
}%

\author{Claudio Toni}

\email{claudio.toni@lapth.cnrs.fr}

\affiliation{%
{\itshape LAPTh, Universit\'{e} Savoie Mont-Blanc et CNRS, 74941 Annecy, France}
}%

\begin{abstract}
\noi
We introduce a model-independent framework to reinterpret Belle~II results using only public data, analytically reconstructing the mapping between true and reconstructed kinematic variables within the statistically dominant Inclusive Tagging Analysis. This enables rare-decay measurements to probe light invisible particles---such as the QCD axion or axion-like particles, collectively denoted $a$---without relying on internal simulations.
Applying the method to $B^+ \! \to \! K^+ \nu \bar\nu$ yields the strongest bound on the branching fraction for $B^+ \! \to \! K^+ a$, improving existing limits by about a factor of nine and constraining the axion's fundamental flavour-changing coupling to $b$ and $s$ quarks. The approach establishes $B^+ \! \to \! K^+ \nu \bar\nu$ as a dual probe---simultaneously testing short-distance new physics and light invisible states, the two probes working independently to an excellent approximation---and provides a general strategy for model-independent reinterpretation of collider data.
\end{abstract}

\maketitle

\bof{Introduction} Axions and axion-like particles (ALPs) are hypothetical light, spinless, pseudoscalar states actively sought through laboratory experiments~\cite{Irastorza:2018dyq,Lanfranchi:2020crw,Sikivie:2020zpn}, astrophysical observations~\cite{Raffelt:1990yz,Raffelt:2006cw,Caputo:2024oqc}, cosmological probes~\cite{Marsh:2015xka}, and theoretical studies~\cite{Kim:2008hd,DiLuzio:2020wdo,Choi:2020rgn}. A central target is the ``invisible'' \cite{Kim:1979if,Shifman:1979if,Dine:1981rt,Zhitnitsky:1980tq} QCD axion~\cite{Peccei:1977hh,Peccei:1977ur,Weinberg:1977ma,Wilczek:1977pj}, henceforth {\em the} axion, which elegantly solves the strong-$CP$ problem---the unexplained suppression of a $CP$-violating QCD term. Given the extreme smallness of this term, $O(10^{-10})$, neither coincidence nor anthropic reasoning offers a satisfactory explanation~\cite{Ubaldi:2008nf,Dine:2018glh}. The axion is also a viable dark-matter candidate~\cite{Preskill:1982cy,Abbott:1982af,Dine:1982ah}. Throughout this paper we denote the QCD axion or a generic ALP as~$a$. Unlike generic ALPs, the QCD-axion's mass $m_a$ and decay constant $f_a$ are linked: a lighter QCD axion necessarily couples more weakly---its interactions scaling as $1/f_a$.

Although this relation places the QCD axion well below the pion mass~\cite{DiLuzio:2020wdo}, it simultaneously opens the door to its production at high-intensity colliders, where large event yields can compensate for its small couplings. 
In this regime, rare decays of heavy mesons such as $K$, $D$, and $B$---or their baryonic counterparts---provide a unique window: the axion would manifest as missing energy and momentum, while the hadronic system, for suitably chosen channels, remains fully reconstructible. 
This defines a clean, general, and conservative search strategy capable of probing both QCD axions and generic ALPs.

A key experimental requirement for such searches is that the visible decay products be measured with high precision and the kinematics of the parent hadron be well constrained, enabling the identification of missing energy with minimal model dependence. 
The Belle II experiment~\cite{Belle-II:2010dht} is particularly well suited in this regard: it collides $e^-$ and $e^+$ beams with momenta of approximately $7$ and $4$~GeV~\footnote{The spread, or bite, around these momenta is tiny and completely negligible for the purposes of this paper.}, producing $b\bar b$ pairs almost exclusively through the $\Upsilon(4S)$ resonance. 
The latter decays with near-unit probability into charged or neutral $B\bar B$ meson pairs. 
Signal decays are then studied in one of the two $B$ mesons, while the other is used as a tag, allowing the kinematics of the signal decay to be reconstructed and the flavour of the signal $B$ to be identified.

In this work we consider the decay $\BpKpa$~\footnote{The charge-conjugate mode is assumed to be included throughout.}.
The axion–hadron interaction relevant for this decay is obtained by matching the axion–quark interaction onto the low-energy chiral effective theory. As a result, the axion–hadron couplings are unambiguously determined---up to the standard hadronic low-energy constants at the chosen chiral order---by the underlying axion–quark couplings. Just below the electroweak scale, the latter can be written in the most general form as~\cite{\georgi}
\begin{align}
\label{eq:Laqq}
\mathcal{L}_a \supset\; &
\frac{\de_\mu a}{2 f_a} \sum_{i,j = 1}^{3}
\bar d_i \gamma^\mu \bigl[ \kV{d_i d_j} + \gamma_5 \kA{d_i d_j} \bigl] d_j
+ \left\{ d \to u \right\} \notag \\
& - \frac{\alpha_s}{8 \pi} \frac{a}{f_a} G \tilde G ~,
\end{align}
where $u_i,d_i$ denote up- and down-type quarks of generation $i$ in the mass eigenbasis. The $G \tilde G$ operator~\footnote{The sign difference in the $f_a$-defining $a G \tilde G$ term in Eq.\ref{eq:Laqq} with respect to Ref.\cite{\diluzio} arises from our use of opposite $\epsilon$-tensor conventions.} can be traded for an axion-field dependent quark mass matrix by redefining the quark fields through $q \to \exp[i (Q_V + Q_A \gamma_5)\, a / (2 f_a)] q$, with $\Tr Q_A = 1$ \cite{\georgi}. Although the hermitian matrices $Q_V, Q_A$ can be chosen as diagonal, they introduce an unphysical reparametrization ambiguity~\cite{\georgi}.

The process $\BpKpa$, on which this paper is focused, involves the complex coupling $\kV{sb}$. The calculation of this process requires the following hadronic matrix element
\be\label{eq:MEs}
\<K^-(p-q)|\bar s b | B^-(p) \> = \frac{m_B^2 - m_K^2}{m_b - m_s} f_0(q^2)~.
\ee
Here, form-factor values and conventions are taken from Ref.~\cite{Gubernari:2023puw}. The relation in \refeq{eq:MEs} implies the following branching ratio (BR)~\cite{\pdg,Bolton:2024egx}
\be\label{eq:BRa}
\mc B(B^- \to K^- a) = \frac{\tau_B |\kV{sb}|^2 \, | \vec k | m_B^2 }{32 \pi f_a^2} 
\left(1 - \frac{m_K^2}{m_B^2}\right)^2 f_0^2(m_a^2)~,
\ee
where $|\vec k| = \lambda^{1/2}(m_B^2,m_{K^{(*)}}^2,m_a^2) / (2 m_B)$ and $\lambda(a,b,c) = a^2+b^2+c^2-2ab-2ac-2bc$. While the theoretically well-motivated QCD axion requires $m_a \simeq 0$, we will also discuss the general case of an ALP. Here and henceforth, the axion-quark couplings are understood at a scale close to $m_B$.

We search for the signal decay $\BpKpa$ in data collected by Belle II for the $\BpKpvv$ measurements. Belle~II provides two nearly {\em uncorrelated} datasets for the $\BpKpvv$ analysis: the ``classic'' Hadronic Tagging Analysis (HTA) and the more recent Inclusive Tagging Analysis (ITA)~\cite{\BelleIIevi,Belle:2019iji,Belle-II:2021rof}. In the HTA, the full tag-$B$ decay chain is reconstructed, allowing the di-neutrino invariant mass $q^2$ to be measured directly. In contrast, the ITA provides far larger statistics---and therefore dominates the sensitivity---but does not reconstruct $q^2$; instead, it uses a proxy variable $q^2_{\mathrm{rec}}$ obtained as discussed below. Recasting the ITA sample for $\BpKpa$ searches therefore requires knowledge of the $q^2$–$q^2_{\mathrm{rec}}$ mapping, which is normally accessible only through internal Belle~II simulation.

A first central result of this work is to show that this mapping can be derived analytically from kinematics alone, without relying on non-public experimental inputs. This insight enables a model-independent reinterpretation of the ITA data for light invisible states such as the axion. Using this mapping and publicly available efficiencies, we demonstrate that this search yields the strongest existing bounds on the magnitudes of the fundamental QCD-axion coupling $\kV{sb}$. This represent our second key result.

Both the ITA and HTA measurements show central values above the Standard-Model (SM) prediction for $B^+ \!\to\! K^+ \nu\bar\nu$~\cite{Parrott:2022zte,Becirevic:2023aov,Buras:2014fpa,Brod:2010hi}. This motivates introducing a signal-strength parameter $\mu_{\nu\bar\nu}$, defined through $d\mathcal{B}_{\nu\nu}(\mu)/dq^2 \equiv \mu_{\nu\bar\nu}\, d\mathcal{B}_{\nu\nu,\mathrm{SM}}/dq^2$. Allowing $\mu_{\nu\bar\nu}$ to float ensures that possible short-distance new-physics effects in $B \!\to\! K\nu\bar\nu$ are not misinterpreted as, or artificially absorbed by, the two-body decay $B \!\to\! K a$. This separation is essential: it renders the analysis sensitive to {\em both} types of new phenomena and establishes $\BpKpvv$ as a dual probe---simultaneously testing short-distance modifications of the $b\!\to\! s\nu\bar\nu$ amplitude and the production of light invisible states $a$. As we will demonstrate, the two effects operate independently to an excellent approximation and this represents our third key result.

To implement this reconstruction explicitly, we recall that within the ITA, Belle II cannot directly reconstruct $q^2 \equiv m_{\nunu}^2 = (p - k)^2$ and thus constructs a proxy variable, $q^2_{\rec}$~\cite{\BelleIIevi}, in terms of the energy of the final-state kaon in the $B\bar{B}$ rest frame (denoted by a star) and of the known $s \equiv (p_{e^-} + p_{e^+})^2 = 4 \, E_B^{*2}$.
In this frame, the signal $B$-meson momentum has a fixed magnitude, $|\vec{p}_B^{\,*}| = 0.3298~\GeV$, and is distributed isotropically. In the $B$ rest frame (double star), the kaon from the two-body decay $B\to K a$ would likewise be monochromatic for fixed $q^2=m_a^2$; however, its observed momentum depends on the relative orientation between the two frames, as discussed next. One can write
\begin{widetext}
\be
\label{eq:q2rec}
\begin{split}
q^2_{\rec}~\equiv~ & \frac{s}{4} + m_K^2 - \sqrt{s} E_K^* 
~=~q^2 + E_B^{*2} - m_B^2 - 2 \, \vec{p}_B^{\,*} \cdot \vec{p}_K^{\,*}(q^2) \\
~=~ & q^2 + E_B^{*2} - m_B^2 - 2 |\vec{p}_B^{\,*}|\gamma^{**} \left( \beta^{**} E_K^{**}(q^2) + |\vec{p}_K^{\,**}(q^2)| \cos \theta^{**} \right)~,
\end{split}
\ee
\end{widetext}
where $|\vec{p}_K^{\,**}| = \lambda^{1/2}(m_B^2,m_K^2,q^2) / (2 m_B)$ and $\beta^{**} = |\vec{p}_B^{\,*}| / E_B^* = 0.06234$ is the boost parameter in the Lorentz transformation from the $B$ rest frame to the $B \bar B$ rest frame.
Finally, $\theta^{**}$ is the polar angle of the kaon momentum in the $B$-meson rest frame relative to the $B$-meson momentum in the $B\bar B$ rest frame.
Importantly, $\theta^{**}$ constitutes the only unmeasured quantity in \refeq{eq:q2rec}. The distribution of $q^2_{\rec}$ for a given $q^2$ can therefore be derived from \refeq{eq:q2rec} with $c^{**}\equiv\cos \theta^{**} \in [-1, 1]$ uniformly distributed.

\refeq{eq:q2rec} makes explicit the aforementioned dependence on the relative orientation between the $^*$ and $^{**}$ frames, showing that its lack of knowledge is structural in the ITA: while the kaon produced in $B\to K a$ is monochromatic in the $B$-meson rest frame, reconstructing this energy ($E_K^*$ in the $q^2_{\rec}$ defining equation) requires knowledge of the full decay kinematics. This includes the angle $\theta^{**}$, which in the ITA is not accessible. The monochromatic kaon energy is thus mapped onto a broadened distribution in $q^2_{\rm rec}$, whose shape is fixed almost entirely by kinematics: the sub-percent-level uncertainties~\cite{BelleIITrackingGroup:2020hpx} smearing the energies and momenta that appear in \refeq{eq:q2rec} have a completely negligible impact on the  $q^2(q^2_{\rec})$ distribution.

The physical branching ratios and the observed yields are then related in the ITA as
\begin{widetext}
\be
\label{eq:dNdq2rec}
 \frac{d N_X}{d q^2_{\rec}} ~=~
 N_B \int dq^2 \int d c^{**} \, \frac{1}{2} \, \delta\Bigl(q^2_{\rec} -  q^2_{\rec}(q^2,c^{**})\Bigl) \, \eps_\text{ITA}(q^2) \, \frac{d \mcB_X}{dq^2}
~\equiv~ N_B \int dq^2 \, f_{q^2_{\rec}}(q^2) \, \eps_\text{ITA}(q^2) \, \frac{d \mcB_X}{dq^2}~,
\ee
\end{widetext}
where $N_B = 387(6) \times 10^6$ \cite{\BelleIIevi} \footnote{The below-2\% $N_B$ error has a negligible impact on our analysis, because $n_{\nunu, \bkg}^i$ have an error approaching 10\%.} is the number of $B$ candidates in the given decay channel $X$, $\mcB$ the corresponding BR, and $\eps_\text{ITA}(q^2)$ the ITA-specific selection efficiency as a function of $q^2$, which can be obtained from Ref.~\cite{\BelleIIevi}. The function $q^2_{\rec}(q^2,c^{**})$ appearing in the delta function is the r.h.s. of \refeq{eq:q2rec}.

\begin{figure}[t]
\begin{center}
\includegraphics[scale=0.45]{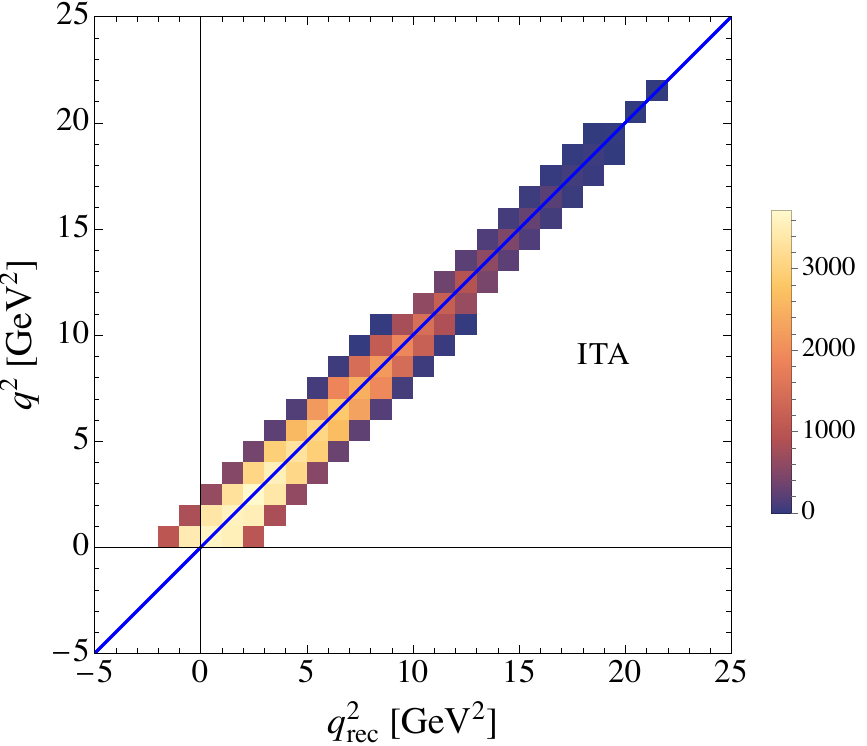}
\caption{Correlation between the true ($q^2$) and the reconstructed ($q^2_{\rec}$) di-neutrino invariant mass squared. The plot is normalised to $10^5$ observed events.}
\label{fig:q2rec_vs_q2}
\end{center}
\end{figure}

In the last equality of \refeq{eq:dNdq2rec} we make contact with the smearing function $f_{q^2_{\rec}}$ \cite{\fridell,\boltonI,\boltonII}, which is typically assumed to require Belle~II simulation. As discussed, this function is instead simply fixed by \refeq{eq:q2rec}, shown in \reffig{fig:q2rec_vs_q2}.
This correlation plot may be compared with Fig.~5.1 of Ref.~\cite{Praz:2022bci}, which is the output of internal Belle II simulations. The two figures display good agreement, aside from an inconsequential low-$q^2$ tail visible in Ref.~\cite{Praz:2022bci}'s figure. This tail originates from radiative decays with undetected photons. Strictly speaking, $B \to K\nu\bar\nu(+n\gamma)$ constitutes a distinct physical process from the non-radiative mode and would need to be subtracted if its contribution were sizeable, as is the case for $B_s \to \mu^+\mu^-$~\cite{Buras:2012ru}. In the present case, however, this effect is negligible---well below the percent level in Ref.~\cite{Praz:2022bci}~\footnote{
The suppression arises because the lightest charged emitter is the $K^+$, which makes the exponent $b$ in
$$
\BR(B \to K\nu\bar \nu(+n\gamma))=\Bigl(\tfrac{2\Delta E}{m_B}\Bigr)^{\tfrac{\alpha}{4\pi} b}\BR(B \to K\nu\bar \nu)
$$
parametrically small. Here $\Delta E$ denotes the minimal photon energy in the $B$ rest frame required for detectability—a ``theoretical’’ scale usually proxied by the $\sqrt{q^2}$ resolution.}. Inclusion of the radiative tail in the referenced figure is therefore immaterial to the above comparison.

\refeq{eq:q2rec} and the ensuing discussion put to work an important distinction between the SM decay $B^+\to K^+\nu\bar\nu$ and the exotic two-body decay $B^+\to K^+ a$: the latter corresponds to a fixed invariant mass $q^2=m_a^2$, whereas the former exhibits a continuous $q^2$ spectrum. This difference is fully exploited in our analysis: it is encoded at the level of the differential branching fractions, which enter as a smooth distribution in $q^2$ for the di-neutrino mode and as a Dirac delta function for the axion signal. The apparent loss of monochromatic information in the ITA does not reflect a limitation of principle, but arises solely from the fact that the experimentally accessible variable is $q^2_{\rm rec}$ rather than the true $q^2$.

By contrast, within the HTA, to which we turn next, Belle~II can reconstruct the full event kinematics, allowing direct access to $q^2$. In this case, the axion signal appears as a narrow peak in the $q^2$ spectrum, equivalent to a monochromatic kaon energy in the $B$-meson rest frame, and the analysis reduces to a standard ``bump search''. The physical BRs and observed yields are simply related as
\be
\label{eq:dNdq2}
 \frac{d N_X}{d q^2} ~=~
 N_B \, \eps_\text{HTA}(q^2) \, \frac{d \mcB_X}{dq^2}~,
\ee
where $\eps_\text{HTA}(q^2)$ is the HTA-specific selection efficiency as a function of $q^2$, also available in Ref.~\cite{\BelleIIevi}.

We conclude by emphasizing that, within our approach, the two tagging strategies exploit the same underlying kinematic information. The observed differences arise solely from degrees of freedom that are not experimentally accessible in the ITA, rather than from any discarded physical information. In particular, the distinction between two-body (signal) and three-body ($B \to K \nu \bar \nu$) kinematics is fully incorporated in both tagging strategies.

\bof{Analysis} To probe the $\BpKpa$ decay through the $\BpKpvv$ measurements at Belle II, we construct a likelihood ratio test statistic based solely on the public data of Fig. 17 (ITA) and Fig. 20 (HTA) of Ref.~\cite{\BelleIIevi}. The likelihood function $\mcL$ is constructed as the product of two contributions, i.e. ${\mcL}={\mcL}_1 {\mcL}_2$.

The first and main term in the likelihood function represents a Poisson distribution $P$ for each bin, labeled by an index $i$, assuming the bins to be uncorrelated: ${\mc L}_1 = \prod_i^{n_\bin} P(n^i_\obs; \lambda^i)$.
Here $n^i_\obs$ is the number of observed events in bin $i$, defined by $q^2 \in [q^2_i, q^2_{i+1})$, and $\lambda^i$ is the corresponding expected number of events, which depends on $\mc B(\BKa) \equiv \mc B_a$ and on the signal strength $\mu_{\nunu}$---the two key quantities to be determined:
\be
\label{eq:lai}
\begin{split}
\lambda^i(\mua, \mu_{\nunu}) ~=~ \mua \, n^i_a + \mu_{\nunu} \, n^i_{\nu\bar\nu} 
+ n^i_\text{bkg}~.
\end{split}
\ee
This expression applies to either the ITA or HTA dataset, where $n^i_a$, $n^i_{\nunu}$, and $n^i_\bkg$ are the yields in the $i^\th$ bin for the signal, the $\BpKpvv$ decay, and the background, respectively. Within the two methods, the first two components on the r.h.s. of \refeq{eq:lai} are defined as:
\begin{align}
\label{eq:nia_ITA}
n^i_{a,\text{ITA}} =&N_B \int\limits_\text{$i^\text{th}$ bin} dq^2_{\rec} \, f_{q^2_{\rec}}(m_a^2) \, \eps_\text{ITA}(m_a^2)~, \\
\label{eq:nivv_ITA}
n^i_{\nu\bar\nu,\text{ITA}} =&N_B \int\limits_\text{$i^\text{th}$ bin} dq^2_{\rec} \int dq^2 \, f_{q^2_{\rec}}(q^2) \, \eps_\text{ITA}(q^2) \, \frac{d\mc B_{\nunu, \SM}}{dq^2}~,\\
\label{eq:nia_HTA}
n^i_{a,\text{HTA}} =&N_B \int\limits_\text{$i^\text{th}$ bin} dq^2 \, \eps_\text{HTA}(m_a^2) \, g_a(m_a^2)~,\\
\label{eq:nivv_HTA}
n^i_{\nu\bar\nu,\text{HTA}} =&N_B \int\limits_\text{$i^\text{th}$ bin} dq^2 \, \eps_\text{HTA}(q^2) \, \frac{d\mc B_{\nunu, \SM}}{dq^2}~.
\end{align}
These expressions can be understood directly from \refeqs{eq:dNdq2rec}{eq:dNdq2}, bearing in mind that, for the signal, the differential branching ratio is proportional to a Dirac delta function with support at $q^2=m_a^2$, and that the overall normalization $\mc B_a$ is factored out in \refeq{eq:lai}.

In addition, for the HTA signal contribution one must account for the finite invariant-mass resolution of the experiment. We implement this by replacing $\delta(q^2-m_a^2)$ with a normal distribution $g_a(q^2)$ centered at $m_a^2$, with variance taken as $\sim(450~\MeV)^2$, comparable to the beam-constrained mass resolution of $B$ mesons reconstructed with the hadronic tag~\cite{Keck:2018lcd}.

This prescription also allows us to comment more generally on detector-induced smearing effects. In principle, the same smearing procedure should be applied to the di-neutrino contribution in the HTA, \refeq{eq:nivv_HTA}. While in the axion case this treatment is essential, in the di-neutrino case the impact is much milder because the $q^2$ spectrum of $B\to K\nu\bar\nu$ is already smooth. Numerically, the smeared and unsmeared distributions differ only marginally~\footnote{The dominant effect is that a few events with small positive $q^2$ migrate to slightly negative reconstructed values, thus shifting between adjacent bins of the Belle~II spectrum.} and the resulting likelihood and axion limits are unchanged within the two-digit precision of our results (see \reftab{tab:bounds}).

In the ITA, the situation is fundamentally different. The axion signal is already broadened at the kinematic level: the monochromatic kaon energy in the $B$ rest frame (or, equivalently, the Dirac-delta structure of the axion $q^2$ distribution) is mapped onto a smooth distribution in the experimentally accessible variable $q^2_{\rec}$. The finite experimental resolution on $E_K^*$ therefore introduces only a subleading additional smearing, whose effect we have verified to be numerically negligible in practice.

We next address the role of uncertainties on the selection-efficiency factors $\eps$ appearing in \refeqs{eq:nia_ITA}{eq:nivv_HTA}. The statistical uncertainties on the efficiencies are at the percent or sub-percent level for both channels, as reported in Ref.~\cite{\BelleIIevi}. The systematic components are, in turn, approximately 3\% in ITA and 16\% in HTA (see Tables I and II of Ref.~\cite{\BelleIIevi}). Since the expected signal yields depend linearly on the efficiencies, these uncertainties propagate directly to the extracted BR limits quoted in \reftab{tab:bounds}.
Because the bounds are largely driven by the ITA sample, the corresponding variation in the ITA and combined results would affect only the third significant digit, whereas the HTA-only limits would be impacted already at the second digit. This is reflected in the precision with which we report the results for the two tagging strategies. In practice, varying the efficiencies within their quoted uncertainties does not modify the bounds at the level of precision displayed in \reftab{tab:bounds}.

We now pause on the treatment of systematic uncertainties affecting the last component of \refeq{eq:lai}, namely the background normalization. In the Belle II analysis, the impact of a given systematic uncertainty is quantified by the reduction in the total uncertainty when that source is removed from the fit, as shown in Tables I and II of Ref.~\cite{\BelleIIevi}. While the individual components cannot be summed in quadrature to reconstruct the total uncertainty, their relative magnitudes provide a useful estimate of their importance. From this comparison, the combined effect of the systematic uncertainties other than the background normalization is at the level of $O(50\%)$ or less relative to the background normalization contribution.

To model this component, we reason as follows. Although the uncertainties on the Monte Carlo generated background events are not explicitly provided, we infer that the relative uncertainty on the total normalisation is respectively $\sigma_\text{ITA}^{\bkg}\approx1\%$ and $\sigma_\text{HTA}^{\bkg}\approx10\%$.
These estimates have been obtained from the $q^2$ distribution and its corresponding pull shown in Figs.~17 and 20 of Ref.~\cite{\BelleIIevi}. One can identify the pull as $|n^i_{\obs} - n^i_{\hst}| / \sqrt{n^i_{\obs} + (\sigma^i_{\hst})^2}$, where the subscripts `obs' and `hst' denote respectively the data and the value of the stacked histograms, and $\sigma^i_{\obs}$ the uncertainty in the data point.
Inverting the pull distribution in terms of $\sigma^i_{\hst}$, we find that the relative uncertainties $\sigma^i_{\hst}/n^i_{\hst}$ are approximately constant across bins. We therefore treat them as a single relative uncertainty $\sigma^{\bkg}$ on the overall background normalization, modeled as a nuisance parameter $\theta$ in $n^i_{\bkg}=(1+\sigma^{\bkg} \, \theta) \, \langle n^i_{\bkg}\rangle$. This can be constrained, independently in the ITA and HTA analyses, by the Gaussian likelihood terms
\be
\label{eq:L2}
{\mcL}_2 = \frac{1}{\sqrt{2\pi}} e^{-\theta^2_{\ITA}/2} \times \frac{1}{\sqrt{2\pi}} e^{-\theta^2_{\HTA}/2}~,
\ee
where the expected value $\langle n^i_\bkg\rangle$ of the number of background events for each bin, $n^i_\bkg$, can be read out from Ref.~\cite{\BelleIIevi}.

A strong test that the above procedure captures the bulk of the systematic uncertainties can be performed by studying our profile likelihood $\mc L$  as a function of the signal strength $\mu_{\nunu}$ in the case where the axion is absent, i.e. where ${\mc B}_a=0$, thus making contact with the counterpart in Ref.~\cite{\BelleIIevi}. The result is shown in \reffig{fig:check} in \refapp{app:plots}, whose two panels show the case of neglecting vs. including the systematic uncertainty on the background normalization.
The figure shows that we obtain results well consistent with the collaboration’s counterparts only when the background normalization uncertainty is included. This indicates that our procedure captures the leading systematic effect.
As a further test that the remaining systematics is negligible, we also calculated the same profile likelihood, with uncertainty inflated by a factor 1.5, as if we were missing a residual $O(50\%)$ component. The resulting profile is essentially identical to the case with non-inflated uncertainty.

\bof{Results} For either the ITA or the HTA method, we then minimize $\chi^2(\mua,\mu_{\nu\bar\nu},\theta) \equiv -2\ln {\mcL}$ and require that $\Delta\chi^2\equiv\chi^2-\chi^2_\text{min}\leq \chi^2(90\%, n_\text{dof})$ \cite{\pdg}. This yields a limit on $\mua$ as a function of $m_a$, with the number of degrees of freedom $n_\dof$ determined according to two alternative hypotheses. Specifically, we may either fix the di-neutrino contribution to its SM value, $\mu_{\nunu}=1$, or allow it to float, which increases $n_\dof$ by one. We then evaluate bounds on $\mc B_a$ using ITA or HTA individually, or their combination. Since the ITA and HTA datasets are uncorrelated to a very good approximation, the combined analysis corresponds to multiplying the individual likelihoods. In this case, we minimize $\chi^2(\mua, \mu_{\nunu}, \theta_{\ITA}, \theta_{\HTA})$.

\renewcommand{\arraystretch}{1.2}
\begin{widetext}
\begin{center}
\begin{table}[h]
\centering
\begin{tabular}{cccc}
\toprule
& \textbf{ITA} & \textbf{HTA} & \textbf{Combined} \\
\midrule
\multirow{3}{*}{$\mu_{\nu\bar\nu}=1$} 
& $n_\text{dof}=2$ & $n_\text{dof}=2$ & $n_\text{dof}=3$ \\
& $\mua < 1.1\times10^{-6}$ & $\mua < 5 \times10^{-6}$ & $\mua < 1.4\times10^{-6}$ \\
& \hspace{5mm} $|\FVsb| > 8.4\times10^{8}$ GeV \hspace{5mm} & \hspace{5mm} $|\FVsb| > 4 \times10^{8}$ GeV \hspace{5mm} & \hspace{5mm} $|\FVsb| > 7.6\times10^{8}$ GeV \hspace{5mm} \\
\midrule
\multirow{3}{*}{$\mu_{\nu\bar\nu}\neq1$} 
& $n_\text{dof}=3$ & $n_\text{dof}=3$ & $n_\text{dof}=4$ \\
& $\mua < 1.0\times10^{-6}$ & $\mua < 6 \times10^{-6}$ & $\mua < 1.2\times10^{-6}$ \\
& $|\FVsb| > 8.9\times10^{8}$ GeV & $|\FVsb| > 4 \times10^{8}$ GeV & $|\FVsb| > 8.0\times10^{8}$ GeV \\
\bottomrule
\end{tabular}
\caption{Upper limits on the axion BR and lower limits on the coupling-rescaled Peccei-Quinn scale at 90\% C.L. for the different cases considered. All limits assume $m_a = 0$. See text for more details on the significant digits quoted for the different tagging strategies.}
\label{tab:bounds}
\end{table}
\end{center}
\end{widetext}

\begin{figure*}[t]
\begin{center}
\includegraphics[scale=0.475]{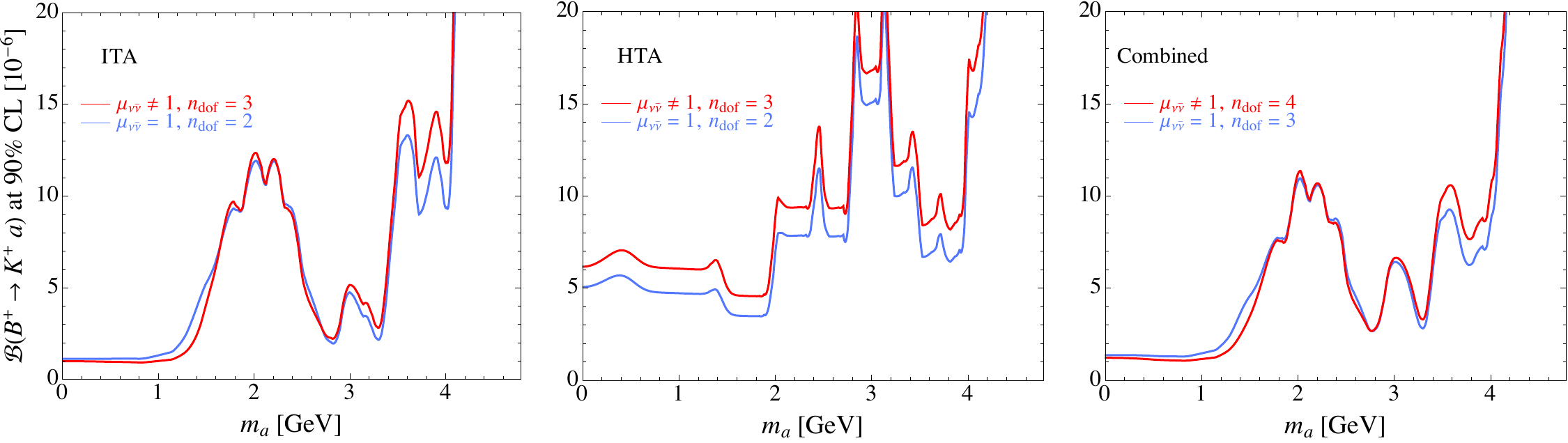}\vspace{0.5cm}
\includegraphics[scale=0.475]{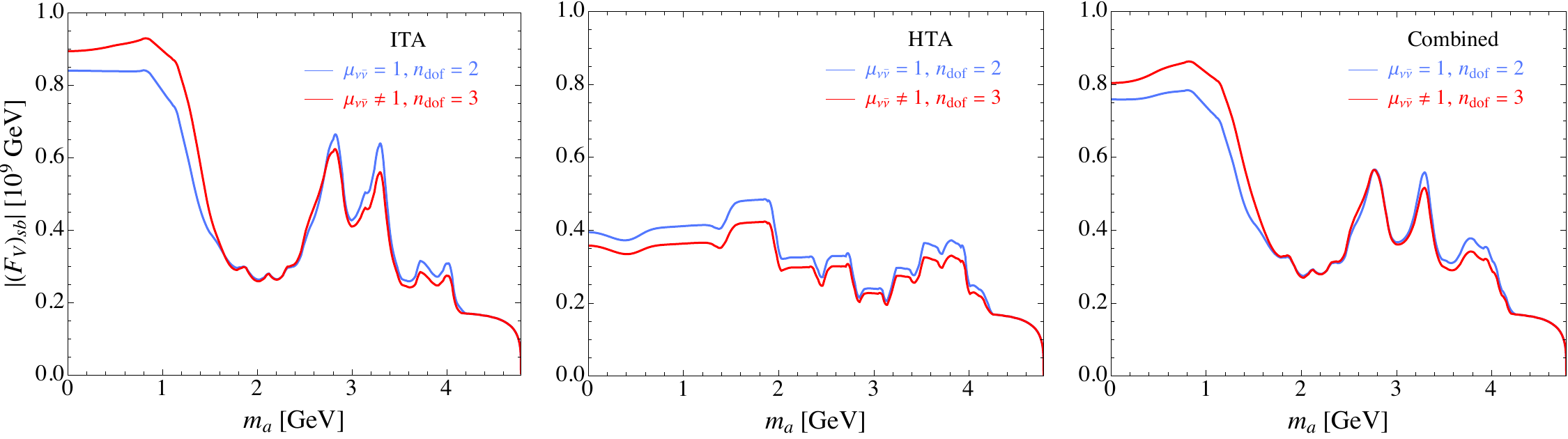}
\caption{Upper limit on $\mcB_a$ (upper row) and lower limit on the coupling-rescaled Peccei-Quinn scale $\FVsb$ (lower row) at 90\% CL as a function of the mass $m_a$, and  with $\mu_{\nunu}$ set to unity (blue line) or left floating (red).}
\label{fig:mass}
\end{center}
\end{figure*}

The different cases described and the resulting bounds are summarized in \reftab{tab:bounds}, which assumes $m_a = 0$~%
\footnote{Ref.~\cite{Ferber:2022rsf} obtains a bound slightly above $1 \times 10^{-6}$ for the branching ratio in the low-$m_a$ regime, based on a projected dataset of 0.5~ab$^{-1}$. Since Ref.~\cite{Ferber:2022rsf} predates Ref.~\cite{\BelleIIevi}, a detailed numerical comparison of the limits is not straightforward. We note however that we achieve a very similar sensitivity---see results in \reftab{tab:bounds}---using 0.362ab$^{-1}$ of actual data, dominated by the ITA selection.}.
Results for a general value of the axion mass are shown in Fig.~\ref{fig:mass}. This figure quantitatively conveys two important pieces of information. First, as one would expect, the bound is weakened by an order of magnitude in the $m_{\nunu}$ bins where data and the SM $\BpKpvv$ prediction are in tension. These discrepant bins can actually be used to set both an upper and a lower bound on $\mcB_a$ \cite{Altmannshofer:2023hkn,\boltonI,\boltonII,Alda:2025uwo}, as shown in \reffig{fig:massBP} in \refapp{app:plots}.

Second, and less obviously, the bound depends only marginally on the assumption of new physics in $\mcB_{\nunu}$---namely, on whether $\mu_{\nunu}$ is set to unity or left floating. This is an important conclusion, making the $\BpKpvv$ decay a {\em double} probe: of new short-distance physics in the $\BpKpvv$ amplitude, testable by $\mu_{\nunu} \neq 1$, and of new light, elusive particles, produced as $B \to K a$, with a BR sensitivity proportional to the total $\mcB_{\nunu}$ uncertainty~\cite{Cavan-Piton:2024pqp}.
This second point is also conveyed by the similarity between the bounds in the first and second rows of \reftab{tab:bounds}. In the case where $\mcB_{\nunu}$ is fixed to its SM value, we verify that varying it by as much as 10\% has a negligible impact on these results.
To corroborate this conclusion even further, we evaluate the bound on $\mcB_a$ assuming fixed values of $\mu_{\nunu}$ but different from unity. As shown in Fig.~\ref{fig:fixed_mununu} of \refapp{app:plots}, the resulting bound has only a mild dependence on the specific value of the di-neutrino signal strength---even for huge beyond-SM contributions to the $\BpKpvv$ amplitude.

\begin{figure}[h]
\begin{center}
\includegraphics[width=0.49\textwidth]{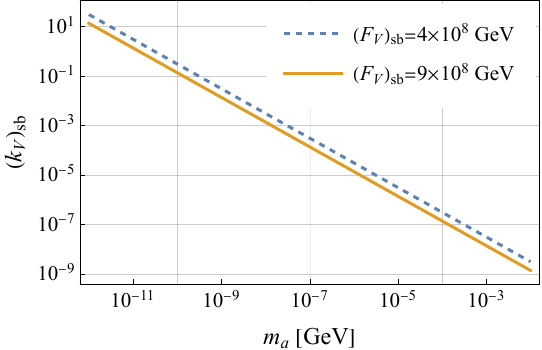}
\caption{QCD axion mass $m_a$ as a function of the effective flavor-violating coupling $(k_V)_{sb}$, shown for two representative lower bounds on $(F_V)_{sb}$ derived in this work. The QCD axion lies effectively at $m_a\simeq 0$ on collider scales, so the resulting limits correspond to vertical constraints in axion mass vs. coupling parameter space.}
\label{fig:kV_ma}
\end{center}
\end{figure}

The BR limit in \reffig{fig:mass} (upper row) can be translated into a bound on the coupling-rescaled Peccei-Quinn scale $\FVsb \equiv 2 f_a / \kV{sb}$ \cite{\camalich}, to obtain a bound on this scale. This bound is shown in \reffig{fig:mass} (lower row), as a function of $m_a$ for the two alternative $\mu_{\nunu}$ possibilities. Limits for $m_a\simeq0$ are collected in Table~\ref{tab:bounds}. As the table shows, we obtain $|\FVsb| \ge 0.9 \times 10^9$ GeV for the QCD axion, which improves by one order of magnitude the pre-existing bound \cite{\camalich}.

We now close the circle with the discussion in the Introduction concerning the distinction between the theoretically motivated but experimentally challenging QCD axion, and generic ALPs, whose appeal lies mostly in their experimental accessibility. With our approach (based on a fully reproducible analysis strategy) we obtain novel bounds on the QCD axion---despite the fact that the QCD axion is generally much more challenging than ALPs at colliders, since its mass scale lies far below any collider ``scales'', as clarified below.

More quantitatively, the axion mass is related to the $(F_V)_{sb}$ scale as
\be
m_a = \frac{\sqrt2 F_\pi m_\pi}{(F_V)_{sb}(k_V)_{sb}} \frac{\sqrt{m_u m_d}}{m_u + m_d}~,
\ee
where $F_\pi = 93 \sqrt2~\MeV$. Our analysis yields lower bounds on $(F_V)_{sb}$, with the weakest at about $4\times10^8$ GeV and the strongest close to $9\times10^8$ GeV, depending on the treatment of the di-neutrino signal strength and on the tagging strategy (hadronic or inclusive) (see \reftab{tab:bounds}). One can therefore study the corresponding range of $m_a$ values as a function of $|(k_V)_{sb}| \lesssim 10$, as imposed by perturbativity. The result is shown in \reffig{fig:kV_ma}.

The figure shows that, over many orders of magnitude in the axion mass, the QCD axion is effectively located at $m_a\simeq 0$ on collider scales---meaning, collectively, the physical effective scales of the processes studied as well as the detection-induced resolution scales. Consequently, our limits act as essentially {\em vertical} constraints in axion parameter spaces of mass vs. coupling, constraining the effective flavor-violating coupling $(k_V)_{sb}$ independently of $m_a$. This representation clarifies how our results should be interpreted within the broader axion-search landscape, which includes experiments that {\em are} sensitive to the axion mass range in our \reffig{fig:kV_ma}: while collider experiments are insensitive to the tiny QCD axion mass itself, they can nevertheless impose meaningful and novel constraints on its flavor structure.

In conclusion, we reinterpret the $\BpKpvv$ measurement as a probe of the two-body decay $\BpKpa$, with $a$ an axion ($m_a \simeq 0$) or a generic ALP ($m_a>0$). A central ingredient of this work is the analytic reconstruction of the mapping between the measured $q^2_{\mathrm{rec}}$ and the true $q^2$, allowing the Inclusive Tag Analysis to be fully used without relying on internal simulations. Combined with publicly available efficiencies, this yields the strongest existing limit on the Peccei–Quinn scale rescaled by the fundamental vector flavour-changing coupling relevant to our channel. A key outcome is that this bound is largely insensitive to assumptions about the di-neutrino signal strength $\mu_{\nu\bar\nu}$. As a result, the decay $\BpKpvv$ acts as a dual probe: it simultaneously constrains short-distance new physics in the $b\!\to\! s\nu\bar\nu$ amplitude and the production of light invisible states such as the axion---with the two effects remaining statistically independent to excellent accuracy. The strategy developed here can be applied directly to $B^+\to K^{*+} \nu \bar\nu$ and, more generally, to other Belle II channels as soon as the corresponding measurements become available, providing a model-independent framework for public-data reinterpretation in rare decays.

\bof{Acknowledgments} We thank Joel C. Swallow for useful discussions.
This work has received funding from the French ANR, under contracts ANR-19-CE31-0016 (`GammaRare') and ANR-23-CE31-0018 (`InvISYble'), that we gratefully acknowledge.

\bibliography{bibliography}

\begin{thebibliography}{53}%
\makeatletter
\providecommand \@ifxundefined [1]{%
 \@ifx{#1\undefined}
}%
\providecommand \@ifnum [1]{%
 \ifnum #1\expandafter \@firstoftwo
 \else \expandafter \@secondoftwo
 \fi
}%
\providecommand \@ifx [1]{%
 \ifx #1\expandafter \@firstoftwo
 \else \expandafter \@secondoftwo
 \fi
}%
\providecommand \natexlab [1]{#1}%
\providecommand \enquote  [1]{``#1''}%
\providecommand \bibnamefont  [1]{#1}%
\providecommand \bibfnamefont [1]{#1}%
\providecommand \citenamefont [1]{#1}%
\providecommand \href@noop [0]{\@secondoftwo}%
\providecommand \href [0]{\begingroup \@sanitize@url \@href}%
\providecommand \@href[1]{\@@startlink{#1}\@@href}%
\providecommand \@@href[1]{\endgroup#1\@@endlink}%
\providecommand \@sanitize@url [0]{\catcode `\\12\catcode `\$12\catcode
  `\&12\catcode `\#12\catcode `\^12\catcode `\_12\catcode `\%12\relax}%
\providecommand \@@startlink[1]{}%
\providecommand \@@endlink[0]{}%
\providecommand \url  [0]{\begingroup\@sanitize@url \@url }%
\providecommand \@url [1]{\endgroup\@href {#1}{\urlprefix }}%
\providecommand \urlprefix  [0]{URL }%
\providecommand \Eprint [0]{\href }%
\providecommand \doibase [0]{https://doi.org/}%
\providecommand \selectlanguage [0]{\@gobble}%
\providecommand \bibinfo  [0]{\@secondoftwo}%
\providecommand \bibfield  [0]{\@secondoftwo}%
\providecommand \translation [1]{[#1]}%
\providecommand \BibitemOpen [0]{}%
\providecommand \bibitemStop [0]{}%
\providecommand \bibitemNoStop [0]{.\EOS\space}%
\providecommand \EOS [0]{\spacefactor3000\relax}%
\providecommand \BibitemShut  [1]{\csname bibitem#1\endcsname}%
\let\auto@bib@innerbib\@empty
\bibitem [{\citenamefont {Irastorza}\ and\ \citenamefont
  {Redondo}(2018)}]{Irastorza:2018dyq}%
  \BibitemOpen
  \bibfield  {author} {\bibinfo {author} {\bibfnamefont {I.~G.}\ \bibnamefont
  {Irastorza}}\ and\ \bibinfo {author} {\bibfnamefont {J.}~\bibnamefont
  {Redondo}},\ }\bibfield  {title} {\bibinfo {title} {{New experimental
  approaches in the search for axion-like particles}},\ }\href
  {https://doi.org/10.1016/j.ppnp.2018.05.003} {\bibfield  {journal} {\bibinfo
  {journal} {Prog. Part. Nucl. Phys.}\ }\textbf {\bibinfo {volume} {102}},\
  \bibinfo {pages} {89} (\bibinfo {year} {2018})},\ \Eprint
  {https://arxiv.org/abs/1801.08127} {arXiv:1801.08127 [hep-ph]} \BibitemShut
  {NoStop}%
\bibitem [{\citenamefont {Lanfranchi}\ \emph {et~al.}(2021)\citenamefont
  {Lanfranchi}, \citenamefont {Pospelov},\ and\ \citenamefont
  {Schuster}}]{Lanfranchi:2020crw}%
  \BibitemOpen
  \bibfield  {author} {\bibinfo {author} {\bibfnamefont {G.}~\bibnamefont
  {Lanfranchi}}, \bibinfo {author} {\bibfnamefont {M.}~\bibnamefont
  {Pospelov}},\ and\ \bibinfo {author} {\bibfnamefont {P.}~\bibnamefont
  {Schuster}},\ }\bibfield  {title} {\bibinfo {title} {{The Search for Feebly
  Interacting Particles}},\ }\href
  {https://doi.org/10.1146/annurev-nucl-102419-055056} {\bibfield  {journal}
  {\bibinfo  {journal} {Ann. Rev. Nucl. Part. Sci.}\ }\textbf {\bibinfo
  {volume} {71}},\ \bibinfo {pages} {279} (\bibinfo {year} {2021})},\ \Eprint
  {https://arxiv.org/abs/2011.02157} {arXiv:2011.02157 [hep-ph]} \BibitemShut
  {NoStop}%
\bibitem [{\citenamefont {Sikivie}(2021)}]{Sikivie:2020zpn}%
  \BibitemOpen
  \bibfield  {author} {\bibinfo {author} {\bibfnamefont {P.}~\bibnamefont
  {Sikivie}},\ }\bibfield  {title} {\bibinfo {title} {{Invisible Axion Search
  Methods}},\ }\href {https://doi.org/10.1103/RevModPhys.93.015004} {\bibfield
  {journal} {\bibinfo  {journal} {Rev. Mod. Phys.}\ }\textbf {\bibinfo {volume}
  {93}},\ \bibinfo {pages} {015004} (\bibinfo {year} {2021})},\ \Eprint
  {https://arxiv.org/abs/2003.02206} {arXiv:2003.02206 [hep-ph]} \BibitemShut
  {NoStop}%
\bibitem [{\citenamefont {Raffelt}(1990)}]{Raffelt:1990yz}%
  \BibitemOpen
  \bibfield  {author} {\bibinfo {author} {\bibfnamefont {G.~G.}\ \bibnamefont
  {Raffelt}},\ }\bibfield  {title} {\bibinfo {title} {{Astrophysical methods to
  constrain axions and other novel particle phenomena}},\ }\href
  {https://doi.org/10.1016/0370-1573(90)90054-6} {\bibfield  {journal}
  {\bibinfo  {journal} {Phys. Rept.}\ }\textbf {\bibinfo {volume} {198}},\
  \bibinfo {pages} {1} (\bibinfo {year} {1990})}\BibitemShut {NoStop}%
\bibitem [{\citenamefont {Raffelt}(2008)}]{Raffelt:2006cw}%
  \BibitemOpen
  \bibfield  {author} {\bibinfo {author} {\bibfnamefont {G.~G.}\ \bibnamefont
  {Raffelt}},\ }\bibfield  {title} {\bibinfo {title} {{Astrophysical axion
  bounds}},\ }\href {https://doi.org/10.1007/978-3-540-73518-2_3} {\bibfield
  {journal} {\bibinfo  {journal} {Lect. Notes Phys.}\ }\textbf {\bibinfo
  {volume} {741}},\ \bibinfo {pages} {51} (\bibinfo {year} {2008})},\ \Eprint
  {https://arxiv.org/abs/hep-ph/0611350} {arXiv:hep-ph/0611350} \BibitemShut
  {NoStop}%
\bibitem [{\citenamefont {Caputo}\ and\ \citenamefont
  {Raffelt}(2024)}]{Caputo:2024oqc}%
  \BibitemOpen
  \bibfield  {author} {\bibinfo {author} {\bibfnamefont {A.}~\bibnamefont
  {Caputo}}\ and\ \bibinfo {author} {\bibfnamefont {G.}~\bibnamefont
  {Raffelt}},\ }\bibfield  {title} {\bibinfo {title} {{Astrophysical Axion
  Bounds: The 2024 Edition}},\ }in\ \href {https://doi.org/10.22323/1.454.0041}
  {\emph {\bibinfo {booktitle} {{1st Training School of the COST Action COSMIC
  WISPers (CA21106)}}}}\ (\bibinfo {year} {2024})\ \Eprint
  {https://arxiv.org/abs/2401.13728} {arXiv:2401.13728 [hep-ph]} \BibitemShut
  {NoStop}%
\bibitem [{\citenamefont {Marsh}(2016)}]{Marsh:2015xka}%
  \BibitemOpen
  \bibfield  {author} {\bibinfo {author} {\bibfnamefont {D.~J.~E.}\
  \bibnamefont {Marsh}},\ }\bibfield  {title} {\bibinfo {title} {{Axion
  Cosmology}},\ }\href {https://doi.org/10.1016/j.physrep.2016.06.005}
  {\bibfield  {journal} {\bibinfo  {journal} {Phys. Rept.}\ }\textbf {\bibinfo
  {volume} {643}},\ \bibinfo {pages} {1} (\bibinfo {year} {2016})},\ \Eprint
  {https://arxiv.org/abs/1510.07633} {arXiv:1510.07633 [astro-ph.CO]}
  \BibitemShut {NoStop}%
\bibitem [{\citenamefont {Kim}\ and\ \citenamefont
  {Carosi}(2010)}]{Kim:2008hd}%
  \BibitemOpen
  \bibfield  {author} {\bibinfo {author} {\bibfnamefont {J.~E.}\ \bibnamefont
  {Kim}}\ and\ \bibinfo {author} {\bibfnamefont {G.}~\bibnamefont {Carosi}},\
  }\bibfield  {title} {\bibinfo {title} {{Axions and the Strong CP Problem}},\
  }\href {https://doi.org/10.1103/RevModPhys.82.557} {\bibfield  {journal}
  {\bibinfo  {journal} {Rev. Mod. Phys.}\ }\textbf {\bibinfo {volume} {82}},\
  \bibinfo {pages} {557} (\bibinfo {year} {2010})},\ \bibinfo {note} {[Erratum:
  Rev.Mod.Phys. 91, 049902 (2019)]},\ \Eprint {https://arxiv.org/abs/0807.3125}
  {arXiv:0807.3125 [hep-ph]} \BibitemShut {NoStop}%
\bibitem [{\citenamefont {Di~Luzio}\ \emph {et~al.}(2020)\citenamefont
  {Di~Luzio}, \citenamefont {Giannotti}, \citenamefont {Nardi},\ and\
  \citenamefont {Visinelli}}]{DiLuzio:2020wdo}%
  \BibitemOpen
  \bibfield  {author} {\bibinfo {author} {\bibfnamefont {L.}~\bibnamefont
  {Di~Luzio}}, \bibinfo {author} {\bibfnamefont {M.}~\bibnamefont {Giannotti}},
  \bibinfo {author} {\bibfnamefont {E.}~\bibnamefont {Nardi}},\ and\ \bibinfo
  {author} {\bibfnamefont {L.}~\bibnamefont {Visinelli}},\ }\bibfield  {title}
  {\bibinfo {title} {{The landscape of QCD axion models}},\ }\href
  {https://doi.org/10.1016/j.physrep.2020.06.002} {\bibfield  {journal}
  {\bibinfo  {journal} {Phys. Rept.}\ }\textbf {\bibinfo {volume} {870}},\
  \bibinfo {pages} {1} (\bibinfo {year} {2020})},\ \Eprint
  {https://arxiv.org/abs/2003.01100} {arXiv:2003.01100 [hep-ph]} \BibitemShut
  {NoStop}%
\bibitem [{\citenamefont {Choi}\ \emph {et~al.}(2021)\citenamefont {Choi},
  \citenamefont {Im},\ and\ \citenamefont {Sub~Shin}}]{Choi:2020rgn}%
  \BibitemOpen
  \bibfield  {author} {\bibinfo {author} {\bibfnamefont {K.}~\bibnamefont
  {Choi}}, \bibinfo {author} {\bibfnamefont {S.~H.}\ \bibnamefont {Im}},\ and\
  \bibinfo {author} {\bibfnamefont {C.}~\bibnamefont {Sub~Shin}},\ }\bibfield
  {title} {\bibinfo {title} {{Recent Progress in the Physics of Axions and
  Axion-Like Particles}},\ }\href
  {https://doi.org/10.1146/annurev-nucl-120720-031147} {\bibfield  {journal}
  {\bibinfo  {journal} {Ann. Rev. Nucl. Part. Sci.}\ }\textbf {\bibinfo
  {volume} {71}},\ \bibinfo {pages} {225} (\bibinfo {year} {2021})},\ \Eprint
  {https://arxiv.org/abs/2012.05029} {arXiv:2012.05029 [hep-ph]} \BibitemShut
  {NoStop}%
\bibitem [{\citenamefont {Kim}(1979)}]{Kim:1979if}%
  \BibitemOpen
  \bibfield  {author} {\bibinfo {author} {\bibfnamefont {J.~E.}\ \bibnamefont
  {Kim}},\ }\bibfield  {title} {\bibinfo {title} {{Weak Interaction Singlet and
  Strong CP Invariance}},\ }\href {https://doi.org/10.1103/PhysRevLett.43.103}
  {\bibfield  {journal} {\bibinfo  {journal} {Phys. Rev. Lett.}\ }\textbf
  {\bibinfo {volume} {43}},\ \bibinfo {pages} {103} (\bibinfo {year}
  {1979})}\BibitemShut {NoStop}%
\bibitem [{\citenamefont {Shifman}\ \emph {et~al.}(1980)\citenamefont
  {Shifman}, \citenamefont {Vainshtein},\ and\ \citenamefont
  {Zakharov}}]{Shifman:1979if}%
  \BibitemOpen
  \bibfield  {author} {\bibinfo {author} {\bibfnamefont {M.~A.}\ \bibnamefont
  {Shifman}}, \bibinfo {author} {\bibfnamefont {A.~I.}\ \bibnamefont
  {Vainshtein}},\ and\ \bibinfo {author} {\bibfnamefont {V.~I.}\ \bibnamefont
  {Zakharov}},\ }\bibfield  {title} {\bibinfo {title} {{Can Confinement Ensure
  Natural CP Invariance of Strong Interactions?}},\ }\href
  {https://doi.org/10.1016/0550-3213(80)90209-6} {\bibfield  {journal}
  {\bibinfo  {journal} {Nucl. Phys. B}\ }\textbf {\bibinfo {volume} {166}},\
  \bibinfo {pages} {493} (\bibinfo {year} {1980})}\BibitemShut {NoStop}%
\bibitem [{\citenamefont {Dine}\ \emph {et~al.}(1981)\citenamefont {Dine},
  \citenamefont {Fischler},\ and\ \citenamefont {Srednicki}}]{Dine:1981rt}%
  \BibitemOpen
  \bibfield  {author} {\bibinfo {author} {\bibfnamefont {M.}~\bibnamefont
  {Dine}}, \bibinfo {author} {\bibfnamefont {W.}~\bibnamefont {Fischler}},\
  and\ \bibinfo {author} {\bibfnamefont {M.}~\bibnamefont {Srednicki}},\
  }\bibfield  {title} {\bibinfo {title} {{A Simple Solution to the Strong CP
  Problem with a Harmless Axion}},\ }\href
  {https://doi.org/10.1016/0370-2693(81)90590-6} {\bibfield  {journal}
  {\bibinfo  {journal} {Phys. Lett. B}\ }\textbf {\bibinfo {volume} {104}},\
  \bibinfo {pages} {199} (\bibinfo {year} {1981})}\BibitemShut {NoStop}%
\bibitem [{\citenamefont {Zhitnitsky}(1980)}]{Zhitnitsky:1980tq}%
  \BibitemOpen
  \bibfield  {author} {\bibinfo {author} {\bibfnamefont {A.~R.}\ \bibnamefont
  {Zhitnitsky}},\ }\bibfield  {title} {\bibinfo {title} {{On Possible
  Suppression of the Axion Hadron Interactions. (In Russian)}},\ }\href@noop {}
  {\bibfield  {journal} {\bibinfo  {journal} {Sov. J. Nucl. Phys.}\ }\textbf
  {\bibinfo {volume} {31}},\ \bibinfo {pages} {260} (\bibinfo {year}
  {1980})}\BibitemShut {NoStop}%
\bibitem [{\citenamefont {Peccei}\ and\ \citenamefont
  {Quinn}(1977{\natexlab{a}})}]{Peccei:1977hh}%
  \BibitemOpen
  \bibfield  {author} {\bibinfo {author} {\bibfnamefont {R.~D.}\ \bibnamefont
  {Peccei}}\ and\ \bibinfo {author} {\bibfnamefont {H.~R.}\ \bibnamefont
  {Quinn}},\ }\bibfield  {title} {\bibinfo {title} {{CP Conservation in the
  Presence of Instantons}},\ }\href
  {https://doi.org/10.1103/PhysRevLett.38.1440} {\bibfield  {journal} {\bibinfo
   {journal} {Phys. Rev. Lett.}\ }\textbf {\bibinfo {volume} {38}},\ \bibinfo
  {pages} {1440} (\bibinfo {year} {1977}{\natexlab{a}})}\BibitemShut {NoStop}%
\bibitem [{\citenamefont {Peccei}\ and\ \citenamefont
  {Quinn}(1977{\natexlab{b}})}]{Peccei:1977ur}%
  \BibitemOpen
  \bibfield  {author} {\bibinfo {author} {\bibfnamefont {R.~D.}\ \bibnamefont
  {Peccei}}\ and\ \bibinfo {author} {\bibfnamefont {H.~R.}\ \bibnamefont
  {Quinn}},\ }\bibfield  {title} {\bibinfo {title} {{Constraints Imposed by CP
  Conservation in the Presence of Instantons}},\ }\href
  {https://doi.org/10.1103/PhysRevD.16.1791} {\bibfield  {journal} {\bibinfo
  {journal} {Phys. Rev. D}\ }\textbf {\bibinfo {volume} {16}},\ \bibinfo
  {pages} {1791} (\bibinfo {year} {1977}{\natexlab{b}})}\BibitemShut {NoStop}%
\bibitem [{\citenamefont {Weinberg}(1978)}]{Weinberg:1977ma}%
  \BibitemOpen
  \bibfield  {author} {\bibinfo {author} {\bibfnamefont {S.}~\bibnamefont
  {Weinberg}},\ }\bibfield  {title} {\bibinfo {title} {{A New Light Boson?}},\
  }\href {https://doi.org/10.1103/PhysRevLett.40.223} {\bibfield  {journal}
  {\bibinfo  {journal} {Phys. Rev. Lett.}\ }\textbf {\bibinfo {volume} {40}},\
  \bibinfo {pages} {223} (\bibinfo {year} {1978})}\BibitemShut {NoStop}%
\bibitem [{\citenamefont {Wilczek}(1978)}]{Wilczek:1977pj}%
  \BibitemOpen
  \bibfield  {author} {\bibinfo {author} {\bibfnamefont {F.}~\bibnamefont
  {Wilczek}},\ }\bibfield  {title} {\bibinfo {title} {{Problem of Strong $P$
  and $T$ Invariance in the Presence of Instantons}},\ }\href
  {https://doi.org/10.1103/PhysRevLett.40.279} {\bibfield  {journal} {\bibinfo
  {journal} {Phys. Rev. Lett.}\ }\textbf {\bibinfo {volume} {40}},\ \bibinfo
  {pages} {279} (\bibinfo {year} {1978})}\BibitemShut {NoStop}%
\bibitem [{\citenamefont {Ubaldi}(2010)}]{Ubaldi:2008nf}%
  \BibitemOpen
  \bibfield  {author} {\bibinfo {author} {\bibfnamefont {L.}~\bibnamefont
  {Ubaldi}},\ }\bibfield  {title} {\bibinfo {title} {{Effects of theta on the
  deuteron binding energy and the triple-alpha process}},\ }\href
  {https://doi.org/10.1103/PhysRevD.81.025011} {\bibfield  {journal} {\bibinfo
  {journal} {Phys. Rev. D}\ }\textbf {\bibinfo {volume} {81}},\ \bibinfo
  {pages} {025011} (\bibinfo {year} {2010})},\ \Eprint
  {https://arxiv.org/abs/0811.1599} {arXiv:0811.1599 [hep-ph]} \BibitemShut
  {NoStop}%
\bibitem [{\citenamefont {Dine}\ \emph {et~al.}(2018)\citenamefont {Dine},
  \citenamefont {Stephenson~Haskins}, \citenamefont {Ubaldi},\ and\
  \citenamefont {Xu}}]{Dine:2018glh}%
  \BibitemOpen
  \bibfield  {author} {\bibinfo {author} {\bibfnamefont {M.}~\bibnamefont
  {Dine}}, \bibinfo {author} {\bibfnamefont {L.}~\bibnamefont
  {Stephenson~Haskins}}, \bibinfo {author} {\bibfnamefont {L.}~\bibnamefont
  {Ubaldi}},\ and\ \bibinfo {author} {\bibfnamefont {D.}~\bibnamefont {Xu}},\
  }\bibfield  {title} {\bibinfo {title} {{Some Remarks on Anthropic Approaches
  to the Strong CP Problem}},\ }\href {https://doi.org/10.1007/JHEP05(2018)171}
  {\bibfield  {journal} {\bibinfo  {journal} {JHEP}\ }\textbf {\bibinfo
  {volume} {05}},\ \bibinfo {pages} {171}},\ \Eprint
  {https://arxiv.org/abs/1801.03466} {arXiv:1801.03466 [hep-th]} \BibitemShut
  {NoStop}%
\bibitem [{\citenamefont {Preskill}\ \emph {et~al.}(1983)\citenamefont
  {Preskill}, \citenamefont {Wise},\ and\ \citenamefont
  {Wilczek}}]{Preskill:1982cy}%
  \BibitemOpen
  \bibfield  {author} {\bibinfo {author} {\bibfnamefont {J.}~\bibnamefont
  {Preskill}}, \bibinfo {author} {\bibfnamefont {M.~B.}\ \bibnamefont {Wise}},\
  and\ \bibinfo {author} {\bibfnamefont {F.}~\bibnamefont {Wilczek}},\
  }\bibfield  {title} {\bibinfo {title} {{Cosmology of the Invisible Axion}},\
  }\href {https://doi.org/10.1016/0370-2693(83)90637-8} {\bibfield  {journal}
  {\bibinfo  {journal} {Phys. Lett. B}\ }\textbf {\bibinfo {volume} {120}},\
  \bibinfo {pages} {127} (\bibinfo {year} {1983})}\BibitemShut {NoStop}%
\bibitem [{\citenamefont {Abbott}\ and\ \citenamefont
  {Sikivie}(1983)}]{Abbott:1982af}%
  \BibitemOpen
  \bibfield  {author} {\bibinfo {author} {\bibfnamefont {L.~F.}\ \bibnamefont
  {Abbott}}\ and\ \bibinfo {author} {\bibfnamefont {P.}~\bibnamefont
  {Sikivie}},\ }\bibfield  {title} {\bibinfo {title} {{A Cosmological Bound on
  the Invisible Axion}},\ }\href {https://doi.org/10.1016/0370-2693(83)90638-X}
  {\bibfield  {journal} {\bibinfo  {journal} {Phys. Lett. B}\ }\textbf
  {\bibinfo {volume} {120}},\ \bibinfo {pages} {133} (\bibinfo {year}
  {1983})}\BibitemShut {NoStop}%
\bibitem [{\citenamefont {Dine}\ and\ \citenamefont
  {Fischler}(1983)}]{Dine:1982ah}%
  \BibitemOpen
  \bibfield  {author} {\bibinfo {author} {\bibfnamefont {M.}~\bibnamefont
  {Dine}}\ and\ \bibinfo {author} {\bibfnamefont {W.}~\bibnamefont
  {Fischler}},\ }\bibfield  {title} {\bibinfo {title} {{The Not So Harmless
  Axion}},\ }\href {https://doi.org/10.1016/0370-2693(83)90639-1} {\bibfield
  {journal} {\bibinfo  {journal} {Phys. Lett. B}\ }\textbf {\bibinfo {volume}
  {120}},\ \bibinfo {pages} {137} (\bibinfo {year} {1983})}\BibitemShut
  {NoStop}%
\bibitem [{\citenamefont {Abe}\ \emph {et~al.}(2010)\citenamefont {Abe} \emph
  {et~al.}}]{Belle-II:2010dht}%
  \BibitemOpen
  \bibfield  {author} {\bibinfo {author} {\bibfnamefont {T.}~\bibnamefont
  {Abe}} \emph {et~al.} (\bibinfo {collaboration} {Belle-II}),\ }\bibfield
  {title} {\bibinfo {title} {{Belle II Technical Design Report}},\ }\href@noop
  {} {\  (\bibinfo {year} {2010})},\ \Eprint {https://arxiv.org/abs/1011.0352}
  {arXiv:1011.0352 [physics.ins-det]} \BibitemShut {NoStop}%
\bibitem [{Note1()}]{Note1}%
  \BibitemOpen
  \bibinfo {note} {The spread, or bite, around these momenta is tiny and
  completely negligible for the purposes of this paper.}\BibitemShut {Stop}%
\bibitem [{Note2()}]{Note2}%
  \BibitemOpen
  \bibinfo {note} {The charge-conjugate mode is assumed to be included
  throughout.}\BibitemShut {Stop}%
\bibitem [{\citenamefont {Georgi}\ \emph {et~al.}(1986)\citenamefont {Georgi},
  \citenamefont {Kaplan},\ and\ \citenamefont {Randall}}]{Georgi:1986df}%
  \BibitemOpen
  \bibfield  {author} {\bibinfo {author} {\bibfnamefont {H.}~\bibnamefont
  {Georgi}}, \bibinfo {author} {\bibfnamefont {D.~B.}\ \bibnamefont {Kaplan}},\
  and\ \bibinfo {author} {\bibfnamefont {L.}~\bibnamefont {Randall}},\
  }\bibfield  {title} {\bibinfo {title} {{Manifesting the Invisible Axion at
  Low-energies}},\ }\href {https://doi.org/10.1016/0370-2693(86)90688-X}
  {\bibfield  {journal} {\bibinfo  {journal} {Phys. Lett. B}\ }\textbf
  {\bibinfo {volume} {169}},\ \bibinfo {pages} {73} (\bibinfo {year}
  {1986})}\BibitemShut {NoStop}%
\bibitem [{Note3()}]{Note3}%
  \BibitemOpen
  \bibinfo {note} {The sign difference in the $f_a$-defining $a G \protect
  \tilde G$ term in Eq.\ref {eq:Laqq} with respect to Ref.\cite
  {DiLuzio:2020wdo} arises from our use of opposite $\epsilon $-tensor
  conventions.}\BibitemShut {Stop}%
\bibitem [{\citenamefont {Gubernari}\ \emph {et~al.}(2023)\citenamefont
  {Gubernari}, \citenamefont {Reboud}, \citenamefont {van Dyk},\ and\
  \citenamefont {Virto}}]{Gubernari:2023puw}%
  \BibitemOpen
  \bibfield  {author} {\bibinfo {author} {\bibfnamefont {N.}~\bibnamefont
  {Gubernari}}, \bibinfo {author} {\bibfnamefont {M.}~\bibnamefont {Reboud}},
  \bibinfo {author} {\bibfnamefont {D.}~\bibnamefont {van Dyk}},\ and\ \bibinfo
  {author} {\bibfnamefont {J.}~\bibnamefont {Virto}},\ }\bibfield  {title}
  {\bibinfo {title} {{Dispersive analysis of $B \to K^{(*)}$ and $B_{s} \to
  \phi$ form factors}},\ }\href {https://doi.org/10.1007/JHEP12(2023)153}
  {\bibfield  {journal} {\bibinfo  {journal} {JHEP}\ }\textbf {\bibinfo
  {volume} {12}},\ \bibinfo {pages} {153}},\ \bibinfo {note} {[Erratum: JHEP
  01, 125 (2025)]},\ \Eprint {https://arxiv.org/abs/2305.06301}
  {arXiv:2305.06301 [hep-ph]} \BibitemShut {NoStop}%
\bibitem [{\citenamefont {Navas}\ \emph {et~al.}(2024)\citenamefont {Navas}
  \emph {et~al.}}]{ParticleDataGroup:2024cfk}%
  \BibitemOpen
  \bibfield  {author} {\bibinfo {author} {\bibfnamefont {S.}~\bibnamefont
  {Navas}} \emph {et~al.} (\bibinfo {collaboration} {Particle Data Group}),\
  }\bibfield  {title} {\bibinfo {title} {{Review of particle physics}},\ }\href
  {https://doi.org/10.1103/PhysRevD.110.030001} {\bibfield  {journal} {\bibinfo
   {journal} {Phys. Rev. D}\ }\textbf {\bibinfo {volume} {110}},\ \bibinfo
  {pages} {030001} (\bibinfo {year} {2024})}\BibitemShut {NoStop}%
\bibitem [{\citenamefont {Bolton}\ \emph {et~al.}(2024)\citenamefont {Bolton},
  \citenamefont {Fajfer}, \citenamefont {Kamenik},\ and\ \citenamefont
  {Novoa-Brunet}}]{Bolton:2024egx}%
  \BibitemOpen
  \bibfield  {author} {\bibinfo {author} {\bibfnamefont {P.~D.}\ \bibnamefont
  {Bolton}}, \bibinfo {author} {\bibfnamefont {S.}~\bibnamefont {Fajfer}},
  \bibinfo {author} {\bibfnamefont {J.~F.}\ \bibnamefont {Kamenik}},\ and\
  \bibinfo {author} {\bibfnamefont {M.}~\bibnamefont {Novoa-Brunet}},\
  }\bibfield  {title} {\bibinfo {title} {{Signatures of light new particles in
  $B\textrightarrow{}K(*) E_{\rm miss}$}},\ }\href
  {https://doi.org/10.1103/PhysRevD.110.055001} {\bibfield  {journal} {\bibinfo
   {journal} {Phys. Rev. D}\ }\textbf {\bibinfo {volume} {110}},\ \bibinfo
  {pages} {055001} (\bibinfo {year} {2024})},\ \bibinfo {note} {[Erratum:
  Phys.Rev.D 111, 039903 (2025)]},\ \Eprint {https://arxiv.org/abs/2403.13887}
  {arXiv:2403.13887 [hep-ph]} \BibitemShut {NoStop}%
\bibitem [{\citenamefont {Adachi}\ \emph {et~al.}(2024)\citenamefont {Adachi}
  \emph {et~al.}}]{Belle-II:2023esi}%
  \BibitemOpen
  \bibfield  {author} {\bibinfo {author} {\bibfnamefont {I.}~\bibnamefont
  {Adachi}} \emph {et~al.} (\bibinfo {collaboration} {Belle-II}),\ }\bibfield
  {title} {\bibinfo {title} {{Evidence for
  B+\textrightarrow{}K+\ensuremath{\nu}\ensuremath{\nu}\textasciimacron{}
  decays}},\ }\href {https://doi.org/10.1103/PhysRevD.109.112006} {\bibfield
  {journal} {\bibinfo  {journal} {Phys. Rev. D}\ }\textbf {\bibinfo {volume}
  {109}},\ \bibinfo {pages} {112006} (\bibinfo {year} {2024})},\ \Eprint
  {https://arxiv.org/abs/2311.14647} {arXiv:2311.14647 [hep-ex]} \BibitemShut
  {NoStop}%
\bibitem [{\citenamefont {Prim}\ \emph {et~al.}(2020)\citenamefont {Prim} \emph
  {et~al.}}]{Belle:2019iji}%
  \BibitemOpen
  \bibfield  {author} {\bibinfo {author} {\bibfnamefont {M.~T.}\ \bibnamefont
  {Prim}} \emph {et~al.} (\bibinfo {collaboration} {Belle}),\ }\bibfield
  {title} {\bibinfo {title} {{Search for $B^+ \to \mu^+\, \nu_\mu$ and $B^+ \to
  \mu^+\, N$ with inclusive tagging}},\ }\href
  {https://doi.org/10.1103/PhysRevD.101.032007} {\bibfield  {journal} {\bibinfo
   {journal} {Phys. Rev. D}\ }\textbf {\bibinfo {volume} {101}},\ \bibinfo
  {pages} {032007} (\bibinfo {year} {2020})},\ \Eprint
  {https://arxiv.org/abs/1911.03186} {arXiv:1911.03186 [hep-ex]} \BibitemShut
  {NoStop}%
\bibitem [{\citenamefont {Abudin\'en}\ \emph {et~al.}(2021)\citenamefont
  {Abudin\'en} \emph {et~al.}}]{Belle-II:2021rof}%
  \BibitemOpen
  \bibfield  {author} {\bibinfo {author} {\bibfnamefont {F.}~\bibnamefont
  {Abudin\'en}} \emph {et~al.} (\bibinfo {collaboration} {Belle-II}),\
  }\bibfield  {title} {\bibinfo {title} {{Search for
  B+\textrightarrow{}K+\ensuremath{\nu}\ensuremath{\nu}\textasciimacron{}
  Decays Using an Inclusive Tagging Method at Belle II}},\ }\href
  {https://doi.org/10.1103/PhysRevLett.127.181802} {\bibfield  {journal}
  {\bibinfo  {journal} {Phys. Rev. Lett.}\ }\textbf {\bibinfo {volume} {127}},\
  \bibinfo {pages} {181802} (\bibinfo {year} {2021})},\ \Eprint
  {https://arxiv.org/abs/2104.12624} {arXiv:2104.12624 [hep-ex]} \BibitemShut
  {NoStop}%
\bibitem [{\citenamefont {Parrott}\ \emph {et~al.}(2023)\citenamefont
  {Parrott}, \citenamefont {Bouchard},\ and\ \citenamefont
  {Davies}}]{Parrott:2022zte}%
  \BibitemOpen
  \bibfield  {author} {\bibinfo {author} {\bibfnamefont {W.~G.}\ \bibnamefont
  {Parrott}}, \bibinfo {author} {\bibfnamefont {C.}~\bibnamefont {Bouchard}},\
  and\ \bibinfo {author} {\bibfnamefont {C.~T.~H.}\ \bibnamefont {Davies}}
  (\bibinfo {collaboration} {HPQCD}),\ }\bibfield  {title} {\bibinfo {title}
  {{Standard Model predictions for
  B\textrightarrow{}K\ensuremath{\ell}+\ensuremath{\ell}-,
  B\textrightarrow{}K\ensuremath{\ell}1-\ensuremath{\ell}2+ and
  B\textrightarrow{}K\ensuremath{\nu}\ensuremath{\nu}\textasciimacron{} using
  form factors from Nf=2+1+1 lattice QCD}},\ }\href
  {https://doi.org/10.1103/PhysRevD.107.014511} {\bibfield  {journal} {\bibinfo
   {journal} {Phys. Rev. D}\ }\textbf {\bibinfo {volume} {107}},\ \bibinfo
  {pages} {014511} (\bibinfo {year} {2023})},\ \bibinfo {note} {[Erratum:
  Phys.Rev.D 107, 119903 (2023)]},\ \Eprint {https://arxiv.org/abs/2207.13371}
  {arXiv:2207.13371 [hep-ph]} \BibitemShut {NoStop}%
\bibitem [{\citenamefont {Be\v{c}irevi\'c}\ \emph {et~al.}(2023)\citenamefont
  {Be\v{c}irevi\'c}, \citenamefont {Piazza},\ and\ \citenamefont
  {Sumensari}}]{Becirevic:2023aov}%
  \BibitemOpen
  \bibfield  {author} {\bibinfo {author} {\bibfnamefont {D.}~\bibnamefont
  {Be\v{c}irevi\'c}}, \bibinfo {author} {\bibfnamefont {G.}~\bibnamefont
  {Piazza}},\ and\ \bibinfo {author} {\bibfnamefont {O.}~\bibnamefont
  {Sumensari}},\ }\bibfield  {title} {\bibinfo {title} {{Revisiting
  $B\rightarrow K^{(*)} \nu {\bar{\nu }}$ decays in the Standard Model and
  beyond}},\ }\href {https://doi.org/10.1140/epjc/s10052-023-11388-z}
  {\bibfield  {journal} {\bibinfo  {journal} {Eur. Phys. J. C}\ }\textbf
  {\bibinfo {volume} {83}},\ \bibinfo {pages} {252} (\bibinfo {year} {2023})},\
  \Eprint {https://arxiv.org/abs/2301.06990} {arXiv:2301.06990 [hep-ph]}
  \BibitemShut {NoStop}%
\bibitem [{\citenamefont {Buras}\ \emph {et~al.}(2015)\citenamefont {Buras},
  \citenamefont {Girrbach-Noe}, \citenamefont {Niehoff},\ and\ \citenamefont
  {Straub}}]{Buras:2014fpa}%
  \BibitemOpen
  \bibfield  {author} {\bibinfo {author} {\bibfnamefont {A.~J.}\ \bibnamefont
  {Buras}}, \bibinfo {author} {\bibfnamefont {J.}~\bibnamefont {Girrbach-Noe}},
  \bibinfo {author} {\bibfnamefont {C.}~\bibnamefont {Niehoff}},\ and\ \bibinfo
  {author} {\bibfnamefont {D.~M.}\ \bibnamefont {Straub}},\ }\bibfield  {title}
  {\bibinfo {title} {{$ B\to {K}^{\left(\ast \right)}\nu \overline{\nu} $
  decays in the Standard Model and beyond}},\ }\href
  {https://doi.org/10.1007/JHEP02(2015)184} {\bibfield  {journal} {\bibinfo
  {journal} {JHEP}\ }\textbf {\bibinfo {volume} {02}},\ \bibinfo {pages}
  {184}},\ \Eprint {https://arxiv.org/abs/1409.4557} {arXiv:1409.4557 [hep-ph]}
  \BibitemShut {NoStop}%
\bibitem [{\citenamefont {Brod}\ \emph {et~al.}(2011)\citenamefont {Brod},
  \citenamefont {Gorbahn},\ and\ \citenamefont {Stamou}}]{Brod:2010hi}%
  \BibitemOpen
  \bibfield  {author} {\bibinfo {author} {\bibfnamefont {J.}~\bibnamefont
  {Brod}}, \bibinfo {author} {\bibfnamefont {M.}~\bibnamefont {Gorbahn}},\ and\
  \bibinfo {author} {\bibfnamefont {E.}~\bibnamefont {Stamou}},\ }\bibfield
  {title} {\bibinfo {title} {{Two-Loop Electroweak Corrections for the $K \to
  \pi \nu \bar{\nu}$ Decays}},\ }\href
  {https://doi.org/10.1103/PhysRevD.83.034030} {\bibfield  {journal} {\bibinfo
  {journal} {Phys. Rev. D}\ }\textbf {\bibinfo {volume} {83}},\ \bibinfo
  {pages} {034030} (\bibinfo {year} {2011})},\ \Eprint
  {https://arxiv.org/abs/1009.0947} {arXiv:1009.0947 [hep-ph]} \BibitemShut
  {NoStop}%
\bibitem [{\citenamefont {Bertacchi}\ \emph {et~al.}(2021)\citenamefont
  {Bertacchi} \emph {et~al.}}]{BelleIITrackingGroup:2020hpx}%
  \BibitemOpen
  \bibfield  {author} {\bibinfo {author} {\bibfnamefont {V.}~\bibnamefont
  {Bertacchi}} \emph {et~al.} (\bibinfo {collaboration} {Belle II Tracking
  Group}),\ }\bibfield  {title} {\bibinfo {title} {{Track finding at Belle
  II}},\ }\href {https://doi.org/10.1016/j.cpc.2020.107610} {\bibfield
  {journal} {\bibinfo  {journal} {Comput. Phys. Commun.}\ }\textbf {\bibinfo
  {volume} {259}},\ \bibinfo {pages} {107610} (\bibinfo {year} {2021})},\
  \Eprint {https://arxiv.org/abs/2003.12466} {arXiv:2003.12466
  [physics.ins-det]} \BibitemShut {NoStop}%
\bibitem [{Note4()}]{Note4}%
  \BibitemOpen
  \bibinfo {note} {The below-2\% $N_B$ error has a negligible impact on our
  analysis, because $n_{\nu \protect \bar \nu , {\protect \rm bkg}}^i$ have an
  error approaching 10\%.}\BibitemShut {Stop}%
\bibitem [{\citenamefont {Fridell}\ \emph {et~al.}(2024)\citenamefont
  {Fridell}, \citenamefont {Ghosh}, \citenamefont {Okui},\ and\ \citenamefont
  {Tobioka}}]{Fridell:2023ssf}%
  \BibitemOpen
  \bibfield  {author} {\bibinfo {author} {\bibfnamefont {K.}~\bibnamefont
  {Fridell}}, \bibinfo {author} {\bibfnamefont {M.}~\bibnamefont {Ghosh}},
  \bibinfo {author} {\bibfnamefont {T.}~\bibnamefont {Okui}},\ and\ \bibinfo
  {author} {\bibfnamefont {K.}~\bibnamefont {Tobioka}},\ }\bibfield  {title}
  {\bibinfo {title} {{Decoding the
  B\textrightarrow{}K\ensuremath{\nu}\ensuremath{\nu} excess at Belle II:
  Kinematics, operators, and masses}},\ }\href
  {https://doi.org/10.1103/PhysRevD.109.115006} {\bibfield  {journal} {\bibinfo
   {journal} {Phys. Rev. D}\ }\textbf {\bibinfo {volume} {109}},\ \bibinfo
  {pages} {115006} (\bibinfo {year} {2024})},\ \Eprint
  {https://arxiv.org/abs/2312.12507} {arXiv:2312.12507 [hep-ph]} \BibitemShut
  {NoStop}%
\bibitem [{\citenamefont {Bolton}\ \emph {et~al.}(2025)\citenamefont {Bolton},
  \citenamefont {Fajfer}, \citenamefont {Kamenik},\ and\ \citenamefont
  {Novoa-Brunet}}]{Bolton:2025fsq}%
  \BibitemOpen
  \bibfield  {author} {\bibinfo {author} {\bibfnamefont {P.~D.}\ \bibnamefont
  {Bolton}}, \bibinfo {author} {\bibfnamefont {S.}~\bibnamefont {Fajfer}},
  \bibinfo {author} {\bibfnamefont {J.~F.}\ \bibnamefont {Kamenik}},\ and\
  \bibinfo {author} {\bibfnamefont {M.}~\bibnamefont {Novoa-Brunet}},\
  }\bibfield  {title} {\bibinfo {title} {{Impact of new invisible particles on
  $B\to K^{(*)} E_{\rm miss}$ observables}},\ }\href@noop {} {\  (\bibinfo
  {year} {2025})},\ \Eprint {https://arxiv.org/abs/2503.19025}
  {arXiv:2503.19025 [hep-ph]} \BibitemShut {NoStop}%
\bibitem [{\citenamefont {Praz}(2022)}]{Praz:2022bci}%
  \BibitemOpen
  \bibfield  {author} {\bibinfo {author} {\bibfnamefont {C.}~\bibnamefont
  {Praz}},\ }\emph {\bibinfo {title} {{Search for B {\textrightarrow}
  K{\ensuremath{\mathit{v}}}{\ensuremath{\mathit{v}}} decays with a machine
  learning method at the Belle II experiment}}},\ \href@noop {} {Ph.D.
  thesis},\ \bibinfo  {school} {U. Hamburg (main), Hamburg U., Hamburg U.,
  Dept. Math.}, \bibinfo {address} {Hamburg} (\bibinfo {year}
  {2022})\BibitemShut {NoStop}%
\bibitem [{\citenamefont {Buras}\ \emph {et~al.}(2012)\citenamefont {Buras},
  \citenamefont {Girrbach}, \citenamefont {Guadagnoli},\ and\ \citenamefont
  {Isidori}}]{Buras:2012ru}%
  \BibitemOpen
  \bibfield  {author} {\bibinfo {author} {\bibfnamefont {A.~J.}\ \bibnamefont
  {Buras}}, \bibinfo {author} {\bibfnamefont {J.}~\bibnamefont {Girrbach}},
  \bibinfo {author} {\bibfnamefont {D.}~\bibnamefont {Guadagnoli}},\ and\
  \bibinfo {author} {\bibfnamefont {G.}~\bibnamefont {Isidori}},\ }\bibfield
  {title} {\bibinfo {title} {{On the Standard Model prediction for BR(B{s,d} to
  mu+ mu-)}},\ }\href {https://doi.org/10.1140/epjc/s10052-012-2172-1}
  {\bibfield  {journal} {\bibinfo  {journal} {Eur. Phys. J. C}\ }\textbf
  {\bibinfo {volume} {72}},\ \bibinfo {pages} {2172} (\bibinfo {year}
  {2012})},\ \Eprint {https://arxiv.org/abs/1208.0934} {arXiv:1208.0934
  [hep-ph]} \BibitemShut {NoStop}%
\bibitem [{Note5()}]{Note5}%
  \BibitemOpen
  \bibinfo {note} {The suppression arises because the lightest charged emitter
  is the $K^+$, which makes the exponent $b$ in $$ {\protect \mathcal B}(B \to
  K\nu \protect \bar \nu (+n\gamma ))=\protect \Bigl (\protect \tfrac {2\Delta
  E}{m_B}\protect \Bigr )^{\protect \tfrac {\alpha }{4\pi } b}{\protect
  \mathcal B}(B \to K\nu \protect \bar \nu ) $$ parametrically small. Here
  $\Delta E$ denotes the minimal photon energy in the $B$ rest frame required
  for detectability—a ``theoretical’’ scale usually proxied by the
  $\protect \sqrt {q^2}$ resolution.}\BibitemShut {Stop}%
\bibitem [{\citenamefont {Keck}\ \emph {et~al.}(2019)\citenamefont {Keck} \emph
  {et~al.}}]{Keck:2018lcd}%
  \BibitemOpen
  \bibfield  {author} {\bibinfo {author} {\bibfnamefont {T.}~\bibnamefont
  {Keck}} \emph {et~al.},\ }\bibfield  {title} {\bibinfo {title} {{The Full
  Event Interpretation}: {An Exclusive Tagging Algorithm for the Belle II
  Experiment}},\ }\href {https://doi.org/10.1007/s41781-019-0021-8} {\bibfield
  {journal} {\bibinfo  {journal} {Comput. Softw. Big Sci.}\ }\textbf {\bibinfo
  {volume} {3}},\ \bibinfo {pages} {6} (\bibinfo {year} {2019})},\ \Eprint
  {https://arxiv.org/abs/1807.08680} {arXiv:1807.08680 [hep-ex]} \BibitemShut
  {NoStop}%
\bibitem [{Note6()}]{Note6}%
  \BibitemOpen
  \bibinfo {note} {The dominant effect is that a few events with small positive
  $q^2$ migrate to slightly negative reconstructed values, thus shifting
  between adjacent bins of the Belle~II spectrum.}\BibitemShut {Stop}%
\bibitem [{Note7()}]{Note7}%
  \BibitemOpen
  \bibinfo {note} {Ref.~\cite {Ferber:2022rsf} obtains a bound slightly above
  $1 \times 10^{-6}$ for the branching ratio in the low-$m_a$ regime, based on
  a projected dataset of 0.5~ab$^{-1}$. Since Ref.~\cite {Ferber:2022rsf}
  predates Ref.~\cite {Belle-II:2023esi}, a detailed numerical comparison of
  the limits is not straightforward. We note however that we achieve a very
  similar sensitivity---see results in Table~\ref {tab:bounds}---using
  0.362ab$^{-1}$ of actual data, dominated by the ITA selection.}\BibitemShut
  {Stop}%
\bibitem [{\citenamefont {Altmannshofer}\ \emph {et~al.}(2024)\citenamefont
  {Altmannshofer}, \citenamefont {Crivellin}, \citenamefont {Haigh},
  \citenamefont {Inguglia},\ and\ \citenamefont
  {Martin~Camalich}}]{Altmannshofer:2023hkn}%
  \BibitemOpen
  \bibfield  {author} {\bibinfo {author} {\bibfnamefont {W.}~\bibnamefont
  {Altmannshofer}}, \bibinfo {author} {\bibfnamefont {A.}~\bibnamefont
  {Crivellin}}, \bibinfo {author} {\bibfnamefont {H.}~\bibnamefont {Haigh}},
  \bibinfo {author} {\bibfnamefont {G.}~\bibnamefont {Inguglia}},\ and\
  \bibinfo {author} {\bibfnamefont {J.}~\bibnamefont {Martin~Camalich}},\
  }\bibfield  {title} {\bibinfo {title} {{Light new physics in
  B{\textrightarrow}K(*){\ensuremath{\nu}}{\ensuremath{\nu}}{\textasciimacron}?}},\
  }\href {https://doi.org/10.1103/PhysRevD.109.075008} {\bibfield  {journal}
  {\bibinfo  {journal} {Phys. Rev. D}\ }\textbf {\bibinfo {volume} {109}},\
  \bibinfo {pages} {075008} (\bibinfo {year} {2024})},\ \Eprint
  {https://arxiv.org/abs/2311.14629} {arXiv:2311.14629 [hep-ph]} \BibitemShut
  {NoStop}%
\bibitem [{\citenamefont {Alda}\ \emph {et~al.}(2025)\citenamefont {Alda},
  \citenamefont {Fuentes~Zamoro}, \citenamefont {Merlo}, \citenamefont
  {Ponce~D{\'\i}az},\ and\ \citenamefont {Rigolin}}]{Alda:2025uwo}%
  \BibitemOpen
  \bibfield  {author} {\bibinfo {author} {\bibfnamefont {J.}~\bibnamefont
  {Alda}}, \bibinfo {author} {\bibfnamefont {M.}~\bibnamefont
  {Fuentes~Zamoro}}, \bibinfo {author} {\bibfnamefont {L.}~\bibnamefont
  {Merlo}}, \bibinfo {author} {\bibfnamefont {X.}~\bibnamefont
  {Ponce~D{\'\i}az}},\ and\ \bibinfo {author} {\bibfnamefont {S.}~\bibnamefont
  {Rigolin}},\ }\bibfield  {title} {\bibinfo {title} {{Comprehensive ALP
  Searches in Meson Decays}},\ }\href@noop {} {\  (\bibinfo {year} {2025})},\
  \Eprint {https://arxiv.org/abs/2507.19578} {arXiv:2507.19578 [hep-ph]}
  \BibitemShut {NoStop}%
\bibitem [{\citenamefont {Cavan-Piton}\ \emph {et~al.}(2024)\citenamefont
  {Cavan-Piton}, \citenamefont {Guadagnoli}, \citenamefont {Iohner},
  \citenamefont {Martinez~Santos},\ and\ \citenamefont
  {Vittorio}}]{Cavan-Piton:2024pqp}%
  \BibitemOpen
  \bibfield  {author} {\bibinfo {author} {\bibfnamefont {M.}~\bibnamefont
  {Cavan-Piton}}, \bibinfo {author} {\bibfnamefont {D.}~\bibnamefont
  {Guadagnoli}}, \bibinfo {author} {\bibfnamefont {A.}~\bibnamefont {Iohner}},
  \bibinfo {author} {\bibfnamefont {D.}~\bibnamefont {Martinez~Santos}},\ and\
  \bibinfo {author} {\bibfnamefont {L.}~\bibnamefont {Vittorio}},\ }\bibfield
  {title} {\bibinfo {title} {{Probing QCD Axions or Axion-like Particles in
  three-body $K$ Decays}},\ }\href@noop {} {\  (\bibinfo {year} {2024})},\
  \Eprint {https://arxiv.org/abs/2411.04170} {arXiv:2411.04170 [hep-ph]}
  \BibitemShut {NoStop}%
\bibitem [{\citenamefont {Martin~Camalich}\ \emph {et~al.}(2020)\citenamefont
  {Martin~Camalich}, \citenamefont {Pospelov}, \citenamefont {Vuong},
  \citenamefont {Ziegler},\ and\ \citenamefont
  {Zupan}}]{MartinCamalich:2020dfe}%
  \BibitemOpen
  \bibfield  {author} {\bibinfo {author} {\bibfnamefont {J.}~\bibnamefont
  {Martin~Camalich}}, \bibinfo {author} {\bibfnamefont {M.}~\bibnamefont
  {Pospelov}}, \bibinfo {author} {\bibfnamefont {P.~N.~H.}\ \bibnamefont
  {Vuong}}, \bibinfo {author} {\bibfnamefont {R.}~\bibnamefont {Ziegler}},\
  and\ \bibinfo {author} {\bibfnamefont {J.}~\bibnamefont {Zupan}},\ }\bibfield
   {title} {\bibinfo {title} {{Quark Flavor Phenomenology of the QCD Axion}},\
  }\href {https://doi.org/10.1103/PhysRevD.102.015023} {\bibfield  {journal}
  {\bibinfo  {journal} {Phys. Rev. D}\ }\textbf {\bibinfo {volume} {102}},\
  \bibinfo {pages} {015023} (\bibinfo {year} {2020})},\ \Eprint
  {https://arxiv.org/abs/2002.04623} {arXiv:2002.04623 [hep-ph]} \BibitemShut
  {NoStop}%
\bibitem [{\citenamefont {Ferber}\ \emph {et~al.}(2023)\citenamefont {Ferber},
  \citenamefont {Filimonova}, \citenamefont {Sch{\"a}fer},\ and\ \citenamefont
  {Westhoff}}]{Ferber:2022rsf}%
  \BibitemOpen
  \bibfield  {author} {\bibinfo {author} {\bibfnamefont {T.}~\bibnamefont
  {Ferber}}, \bibinfo {author} {\bibfnamefont {A.}~\bibnamefont {Filimonova}},
  \bibinfo {author} {\bibfnamefont {R.}~\bibnamefont {Sch{\"a}fer}},\ and\
  \bibinfo {author} {\bibfnamefont {S.}~\bibnamefont {Westhoff}},\ }\bibfield
  {title} {\bibinfo {title} {{Displaced or invisible? ALPs from B decays at
  Belle II}},\ }\href {https://doi.org/10.1007/JHEP04(2023)131} {\bibfield
  {journal} {\bibinfo  {journal} {JHEP}\ }\textbf {\bibinfo {volume} {04}},\
  \bibinfo {pages} {131}},\ \Eprint {https://arxiv.org/abs/2201.06580}
  {arXiv:2201.06580 [hep-ph]} \BibitemShut {NoStop}%
\end{thebibliography}%

\newpage
\onecolumngrid

\newpage

\appendix

\renewcommand{\thesubsection}{\thesection\arabic{subsection}}

\section{Supplementary plots}\label{app:plots}

\noi In this part we collect additional plots which are not necessary to understand the results in the main text, but further corroborate them, or can be used to perform comparisons.

\begin{figure}[t]
\begin{center}
\includegraphics[width=0.99\textwidth]{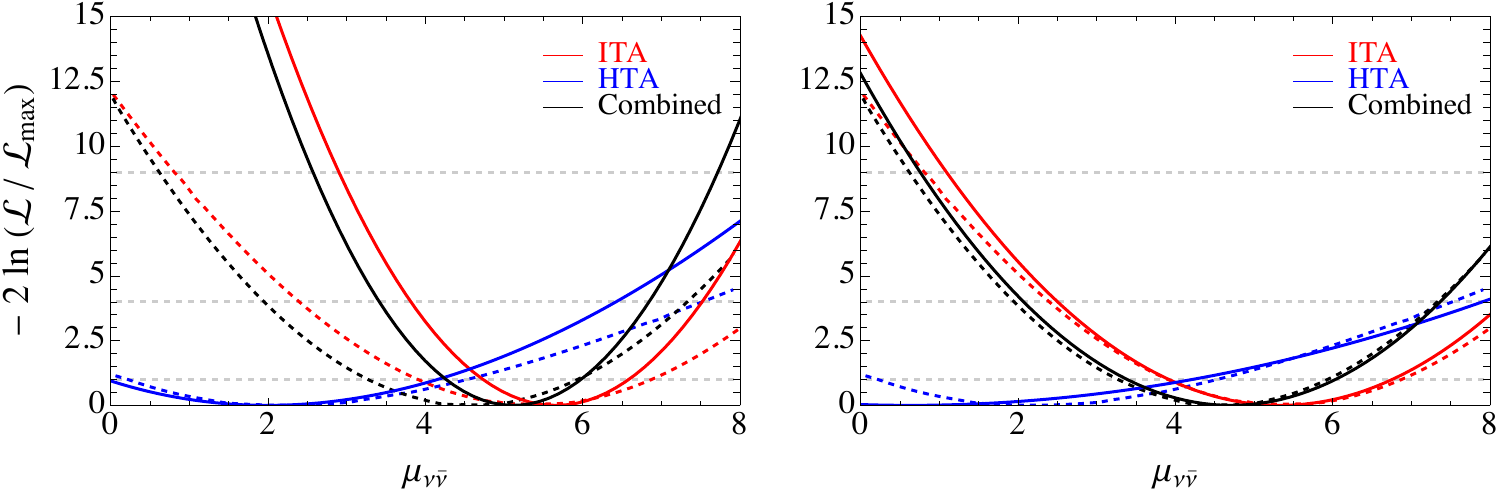}
\caption{Twice the negative profile log-likelihood ratio as a function of $\mu_{\nunu}$ for the ITA (red), HTA (blue), and combined (black) analyses. Solid lines correspond to our results, while dashed lines reproduce the Belle~II results shown in Fig.~16 of Ref.~\cite{\BelleIIevi}. The horizontal grey dashed lines indicate the $1,\,2,\,3\sigma \chi^2$ thresholds for one degree of freedom. (Left) The number of background events in each bin is fixed to its expected value $\langle n^i_{\bkg} \rangle$. (Right) The background normalization is floated as described in the text.}
\label{fig:check}
\end{center}
\end{figure}

\reffig{fig:check} shows the log-likelihood profiles resulting from our analyses and their comparison with the Belle~II counterparts. As the figure indicates, we obtain results very close to those of the collaboration, except in the region $\mu_{\nunu}<1$ for the HTA case. This small deviation is understood: it originates from our treatment of the background uncertainty. When the background is fixed to its central value with no associated error, the minimum coincides with the Belle~II result; including the background error, however, broadens the minimum and slightly lowers the best-fit point.

\begin{figure*}[h]
\begin{center}
\includegraphics[scale=0.45]{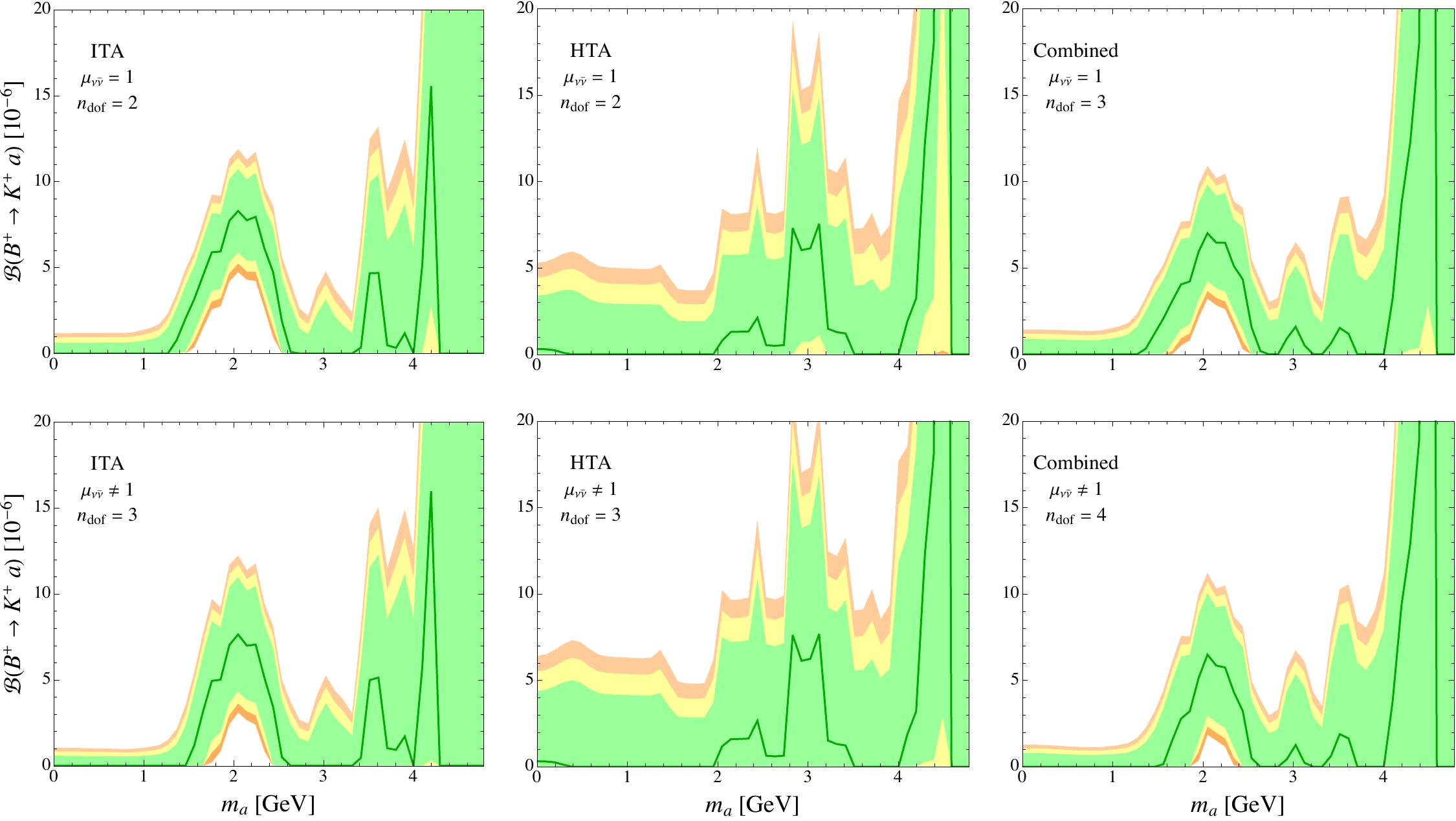}
\caption{Regions $\mcB_a$ at 68\% (green), 95\% (yellow) and 99\% (orange) CL and best fit value (solid green line) of $\mcB_a$ as a function of the mass with $\mu_{\nunu}$ fixed to 1 (upper row) or left floating (lower row).}
\label{fig:massBP}
\end{center}
\end{figure*}

\reffig{fig:massBP} shows $\mcB_a$ as a signal, displaying the $1\sigma$, $2\sigma$, and $3\sigma$ regions for $\mcB_a$ as a function of the mass, with $\mu_{\nunu}$ either fixed to unity or left floating.

\begin{figure*}[h]
\begin{center}
\includegraphics[scale=0.475]{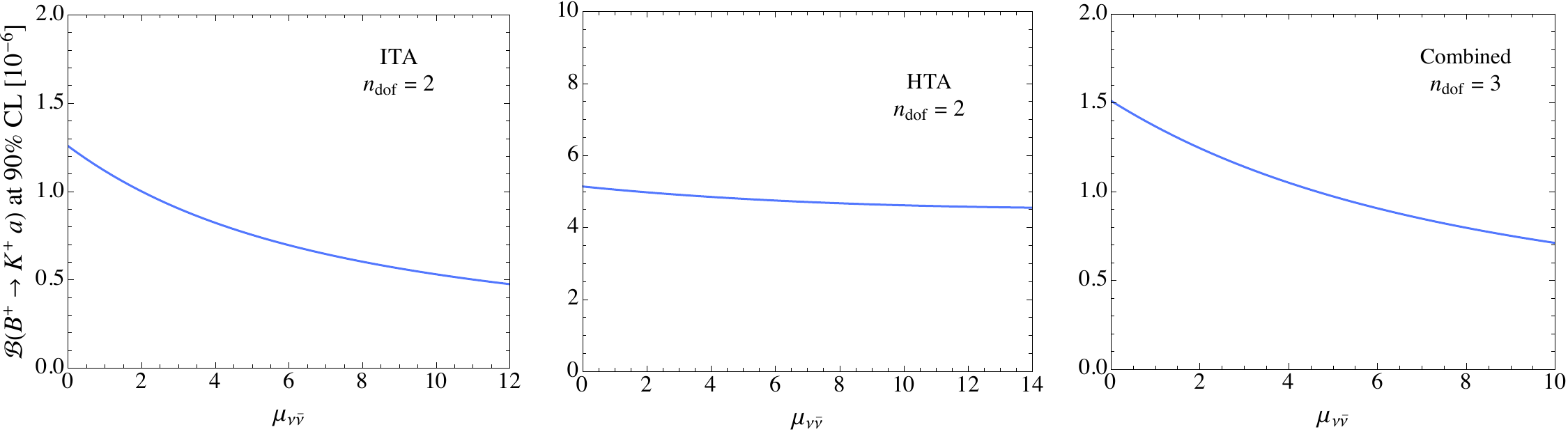}
\caption{Upper limit at 90\% CL on $\mcB_a$ for fixed values of $\mu_{\nunu}$.}
\label{fig:fixed_mununu}
\end{center}
\end{figure*}

In \reffig{fig:fixed_mununu}, we evaluate the bound on $\mcB_a$ for fixed values of $\mu_{\nunu}$ different from unity, illustrating the mild dependence of the resulting bound on the assumed value of the di-neutrino signal strength.

\end{document}